\documentclass[a4paper,11pt]{article}
\begin{document}

\author{E. M. Cioroianu\thanks{e-mail address:
manache@central.ucv.ro} , S. C. S\u {a}raru\thanks{e-mail
address: scsararu@central.ucv.ro} \\
Faculty of Physics, University of Craiova\\
13 A. I. Cuza Str., Craiova, 200585, Romania}
\title{PT-symmetry breaking hamiltonian interactions in BF models}
\date{}
\maketitle

\begin{abstract}
The PT-symmetry breaking, consistent hamiltonian interactions in
all $n\geq 4$ spacetime dimensions that can be added to an abelian
BF model involving a set of scalar fields, two sorts of one-forms,
and a system of two-forms are obtained by means of the hamiltonian
deformation procedure based on local BRST cohomology. This paper
enhances one of our previous works, where only PT-invariant
deformations were considered. The associated coupled theory is an
interacting, topological BF model exhibiting an open gauge algebra
and on-shell reducibility relations.

\end{abstract}

\section{Introduction}

The great advantage of the hamiltonian BRST symmetry \cite{2,8} is
represented by its proper implementation in quantum mechanics \cite{2}
(Chapter 14), and also by an appropriate correlation with the canonical
quantization methods \cite{9}. The understanding of this symmetry from a
cohomological point of view made possible a unitary approach to many
problems in gauge field theory, such as the hamiltonian analysis of
anomalies \cite{10}, the precise relation between local lagrangian and
hamiltonian BRST cohomologies \cite{11}, and, recently, the problem of
obtaining consistent hamiltonian interactions in gauge theories by means of
the deformation theory \cite{appb,12,noi,mpla,ijmpa}.

In this paper we investigate the PT-symmetry breaking, consistent
hamiltonian deformations in any spacetime dimension $n\geq 4$ of a
free abelian topological field theory of BF-type \cite{13}
involving a set of scalar fields, two collections of one-forms,
and a system of two-forms. Actually, this work enhances our
previous results from \cite{noi}, where the interactions were
imposed to preserve PT invariance. Here, we relax this condition
and show that the resulting interactions are accurately described
by a topological field theory with an open algebra of first-class
constraints, that can be interpreted in terms of a Poisson
structure present in various models of two-dimensional gravity
\cite{grav1,grav2,grav3,grav4}. [The analysis of Poisson Sigma
Models, including their relationship to two-dimensional gravity
and the study of classical solutions, can be found in \cite
{psm1,stroblspec,psm2,psm3,psm4,psmn} (see also \cite{ikeda12})].

The plan of the paper is the following. Section 2 briefly reviews
the problem of constructing consistent hamiltonian interactions in
the framework of the BRST formalism, which reduces to solving two
towers of equations that describe the deformation of the BRST
charge, respectively, of the BRST-invariant Hamiltonian associated
with a given ``free'' first-class theory at various orders in the
coupling constant. In Section 3 we determine the hamiltonian BRST
symmetry ($s$) of the free topological theory under study in
$n\geq 4$ spacetime dimensions, which splits as the sum between
the Koszul-Tate differential and the exterior derivative along the
gauge orbits. This model is abelian and $\left( n-2\right) $-stage
reducible, the reducibility relations holding off-shell
(everywhere in the phase space). Next, we solve the main equations
governing the hamiltonian deformation procedure on behalf of the
BRST cohomology of the free theory. In Section 4 we initially
compute, using specific cohomological techniques, the first-order
deformation of the BRST charge, which lies in the cohomological
space of $s$ modulo the spatial part of the exterior spacetime
derivative ($\tilde{d}$) in ghost number one, $H^{1}\left(
s|\tilde{d}\right) $. The first-order deformation of the BRST
charge stops at antighost number $\left( n-1\right) $ and contains
two types of solutions: one that preserves the PT invariance and
is described by two sorts of arbitrary functions involving only
the undifferentiated scalar fields, previously investigated in
\cite{noi}, and the other that breaks the PT invariance and has
not been considered in the literature so far. The latter
deformation is parametrized by a completely antisymmetric `tensor'
of rank $n$ that involves only the undifferentiated scalar fields.
The consistency of the first-order deformation imposes certain
restrictions on these three types of functions depending only on
the undifferentiated scalar fields and allows them to be
parametrized in terms of a single `two-tensor' (in the collection
indices) depending on the scalar fields, that must be
antisymmetric and fulfills a certain identity. The other two
functions are obtained from the derivatives of this `two-tensor'
with respect to the scalar fields. Under these conditions, all the
other deformations, of order two and higher, can be taken to
vanish, and thus the BRST charge of the interacting model that is
consistent to all orders in the deformation parameter is fully
output. Section 5 solves the problem of generating the deformed
BRST-invariant Hamiltonian, which can be taken nonzero only at the
first order in the coupling constant. With the help of these
deformed hamiltonian BRST quantities, in Section 6 we identify the
interacting gauge theory, which is again topological and displays
an open algebra of constraints (the Dirac brackets among the
deformed first-class constraint functions only close on the
first-class constraint surface). The deformed first-class
constraints are of course reducible, but the reducibility
relations hold on-shell (on the first-class constraint surface).
It is interesting to observe that the relaxation of the condition
on the PT invariance of the deformations brings in new,
consistent, nontrivial terms at the level of both BRST charge and
BRST-invariant Hamiltonian. The only sector that `does not feel'
the relaxation of this condition is the redundancy of the deformed
first-class constraints, including both the reducibility relations
and functions. Section 7 contains the main conclusions of the
present paper. Two appendix sections complete the description of
the interacting model.

\section{Main equations of the hamiltonian deformation procedure}

It has been shown in \cite{appb} that the problem of constructing
consistent hamiltonian interactions in theories with first-class
constraints can be equivalently reformulated as a deformation
problem of the BRST charge $\Omega _{0}$ and of the BRST-invariant
Hamiltonian $H_{0\mathrm{B}}$ of a given ``free'' first-class
theory. More precisely, if the interactions can be consistently
constructed, then the ``free'' BRST charge can be deformed into
\begin{eqnarray}
\Omega _{0}\rightarrow \hat{\Omega} &=&\Omega _{0}+g\int d^{n-1}x\,\hat{%
\omega}_{1}+g^{2}\int d^{n-1}x\,\hat{\omega}_{2}+O\left( g^{3}\right) =
\nonumber \\
&=&\Omega _{0}+g\hat{\Omega}_{1}+g^{2}\hat{\Omega}_{2}+O\left( g^{3}\right) ,
\label{d1}
\end{eqnarray}
where the BRST charge of the interacting theory $\hat{\Omega}$ must satisfy
the equation
\begin{equation}
\left[ \hat{\Omega},\hat{\Omega}\right] =0.  \label{d2}
\end{equation}
The last relation projected on various powers in the deformation
parameter $g$ is equivalent with the tower of equations
\begin{eqnarray}
\left[ \Omega _{0},\Omega _{0}\right] &=&0,  \label{d3} \\
2\left[ \Omega _{0},\hat{\Omega}_{1}\right] &=&0,  \label{d4} \\
2\left[ \Omega _{0},\hat{\Omega}_{2}\right] +\left[ \hat{\Omega}_{1},\hat{%
\Omega}_{1}\right] &=&0,  \label{d5} \\
&&\vdots  \nonumber
\end{eqnarray}
In a similar manner the BRST-invariant Hamiltonian of the ``free'' theory
can be deformed like
\begin{eqnarray}
H_{0\mathrm{B}}\rightarrow \hat{H}_{\mathrm{B}} &=&H_{0\mathrm{B}}+g\int
d^{n-1}x\,\hat{h}_{1}+g^{2}\int d^{n-1}x\,\hat{h}_{2}+O\left( g^{3}\right) =
\nonumber \\
&=&H_{0\mathrm{B}}+g\hat{H}_{1}+g^{2}\hat{H}_{2}+O\left( g^{3}\right) ,
\label{d6}
\end{eqnarray}
and it stands for the BRST-invariant Hamiltonian of the coupled system
\begin{equation}
\left[ \hat{H}_{\mathrm{B}},\hat{\Omega}\right] =0.  \label{d7}
\end{equation}
The decomposition of the relation (\ref{d7}) according to the various orders
in the coupling constant reveals a new tower of equations
\begin{eqnarray}
\left[ H_{0\mathrm{B}},\Omega _{0}\right] &=&0,  \label{d8} \\
\left[ \hat{H}_{1},\Omega _{0}\right] +\left[ H_{0\mathrm{B}},\hat{\Omega}%
_{1}\right] &=&0,  \label{d9} \\
\left[ \hat{H}_{2},\Omega _{0}\right] +\left[ \hat{H}_{1},\hat{\Omega}%
_{1}\right] +\left[ H_{0\mathrm{B}},\hat{\Omega}_{2}\right] &=&0,
\label{d10} \\
&&\vdots  \nonumber
\end{eqnarray}
While, the equations (\ref{d3}) and (\ref{d8}) are satisfied since $\Omega
_{0}$ and $H_{0\mathrm{B}}$ are by hypothesis the BRST charge and
respectively the BRST-invariant Hamiltonian of the ``free'' theory, the
resolution of the remaining equations ((\ref{d4})--(\ref{d5}), etc., and (%
\ref{d9})--(\ref{d10}), etc.) by means of cohomological techniques provides
the hamiltonian BRST description of the interacting gauge theory
corresponding to the initial ``free'' one.

\section{Free BRST symmetry\label{free}}

Our starting point is a free, topological field theory of BF-type in $n\geq
4 $ spacetime dimensions that involves two types of one-forms, a collection
of scalar fields, and a system of two-forms, described by the lagrangian
action
\begin{equation}
S_{0}\left[ A_{\mu }^{a},H_{\mu }^{a},\varphi _{a},B_{a}^{\mu \nu }\right]
=\int d^{n}x\left( H_{\mu }^{a}\partial ^{\mu }\varphi _{a}+\frac{1}{2}%
B_{a}^{\mu \nu }\partial _{[\mu }A_{\nu ]}^{a}\right) ,  \label{f1}
\end{equation}
where here and in the sequel the notation $[\mu \ldots \nu ]$ (or $[i\ldots
j]$) signifies full antisymmetry with respect to the indices between
brackets without normalization factors (i.e. the independent terms appear
only once and are not multiplied by overall numerical factors). The above
action is invariant under the gauge transformations
\begin{equation}
\delta _{\epsilon }A_{\mu }^{a}=\partial _{\mu }\epsilon ^{a},\quad \delta
_{\epsilon }H_{\mu }^{a}=\partial ^{\nu }\epsilon _{\mu \nu }^{a},\quad
\delta _{\epsilon }\varphi _{a}=0,\quad \delta _{\epsilon }B_{a}^{\mu \nu
}=\partial _{\rho }\epsilon _{a}^{\mu \nu \rho },  \label{f2}
\end{equation}
which are off-shell $\left( n-2\right) $-stage reducible, where
the gauge parameters $\epsilon ^{a}$, $\epsilon _{\mu \nu }^{a}$,
and $\epsilon _{a}^{\mu \nu \rho }$ are bosonic, the last two sets
being completely antisymmetric.

After the elimination of the second-class constraints (the coordinates of
the reduced phase-space are $z^{A}=\left( \pi _{a}^{0},A_{\mu
}^{a},B_{a}^{\mu \nu },p_{a}^{i},H_{\mu }^{a},\pi _{ij}^{a},\varphi
_{a}\right) $), we are left with a system subject only to the first-class
constraints
\begin{eqnarray}
G_{a}^{(1)} &\equiv &\pi _{a}^{0}\approx 0,\quad G_{a}^{(2)}\equiv -\partial
_{i}B_{a}^{0i}\approx 0,  \label{f3} \\
G_{ij}^{(1)a} &\equiv &2\pi _{ij}^{a}\approx 0,\quad G_{ij}^{(2)a}\equiv
-\partial _{[i}A_{j]}^{a}\approx 0,  \label{f4} \\
\gamma _{a}^{(1)i} &\equiv &-p_{a}^{i}\approx 0,\quad \gamma
_{a}^{(2)i}\equiv \partial ^{i}\varphi _{a}\approx 0,  \label{f5}
\end{eqnarray}
and displaying the first-class Hamiltonian
\begin{equation}
H_{0}=\int d^{n-1}x\left( -H_{i}^{a}\gamma _{a}^{(2)i}+\frac{1}{2}%
B_{a}^{ij}G_{ij}^{(2)a}+A_{0}^{a}G_{a}^{(2)}\right) ,  \label{f6}
\end{equation}
in terms of the non-vanishing fundamental Dirac brackets
\begin{eqnarray}
\left[ \pi _{a}^{0}(t,\mathbf{x}),A_{0}^{b}(t,\mathbf{y})\right] &=&-\delta
_{a}^{b}\delta ^{n-1}\left( \mathbf{x-y}\right) ,  \label{f7} \\
\left[ B_{a}^{0i}(t,\mathbf{x}),A_{j}^{b}(t,\mathbf{y})\right] &=&-\delta
_{j}^{i}\delta _{a}^{b}\delta ^{n-1}\left( \mathbf{x-y}\right) ,  \label{f8}
\\
\left[ H_{0}^{a}(t,\mathbf{x}),\varphi _{b}(t,\mathbf{y})\right] &=&-\delta
_{b}^{a}\delta ^{n-1}\left( \mathbf{x-y}\right) ,  \label{f9} \\
\left[ \pi _{ij}^{a}(t,\mathbf{x}),B_{b}^{kl}(t,\mathbf{y})\right] &=&-\frac{%
1}{2}\delta _{i}^{[k}\delta _{j}^{l]}\delta _{b}^{a}\delta ^{n-1}\left(
\mathbf{x-y}\right) ,  \label{f10} \\
\left[ p_{a}^{i}(t,\mathbf{x}),H_{j}^{b}(t,\mathbf{y})\right] &=&-\delta
_{j}^{i}\delta _{a}^{b}\delta ^{n-1}\left( \mathbf{x-y}\right) .  \label{f11}
\end{eqnarray}
The above constraints are abelian, while the remaining gauge algebra
relations are expressed by
\begin{eqnarray}
\left[ H_{0},G_{a}^{(1)}\right] &=&G_{a}^{(2)},\quad \left[
H_{0},G_{a}^{(2)}\right] =0,  \label{f12} \\
\left[ H_{0},G_{ij}^{(1)a}\right] &=&G_{ij}^{(2)a},\quad \left[
H_{0},G_{ij}^{(2)a}\right] =0,  \label{f13} \\
\left[ H_{0},\gamma _{a}^{(1)i}\right] &=&\gamma _{a}^{(2)i},\quad \left[
H_{0},\gamma _{a}^{(2)i}\right] =0.  \label{f14}
\end{eqnarray}
The constraint functions $G_{ij}^{(2)a}$ are off-shell $\left( n-3\right) $%
-stage reducible, with the reducibility functions (of order $(k-2)$) given
by
\begin{equation}
\left( Z_{i_{1}i_{2}\ldots i_{k}}^{a}\right) _{b}^{j_{1}\ldots j_{k-1}}=%
\frac{\left( -\right) ^{k-1}}{\left( k-1\right) !}\delta _{b}^{a}\partial
_{[i_{1}}\delta _{i_{2}}^{j_{1}}\cdots \delta _{i_{k}]}^{j_{k-1}},\quad k=%
\overline{3,n-1},  \label{f15}
\end{equation}
while the constraint functions $\gamma _{a}^{(2)i}$ are off-shell $\left(
n-2\right) $-stage reducible, the associated reducibility functions (of
order $(k-1)$) being
\begin{equation}
\left( Z_{a}^{i_{1}i_{2}\ldots i_{k}}\right) _{j_{1}\ldots j_{k-1}}^{b}=%
\frac{\left( -\right) ^{k-1}}{\left( k-1\right) !}\delta _{a}^{b}\partial
^{[i_{1}}\delta _{j_{1}}^{i_{2}}\cdots \delta _{j_{k-1}}^{i_{k}]},\quad k=%
\overline{2,n-1}.  \label{f16}
\end{equation}

The hamiltonian BRST formalism requires the introduction of the ghosts
\begin{eqnarray}
\eta ^{a_{0}} &=&\left( \eta ^{(1)a},\eta ^{a},\eta _{a}^{(1)ij},\eta
_{a}^{ij},C_{i}^{(1)a},C_{i}^{a}\right) ,  \label{f17} \\
\eta ^{a_{k}} &=&\left( C_{i_{1}\ldots i_{k+1}}^{a},\eta _{a}^{i_{1}\ldots
i_{k+2}}\right) ,\quad k=\overline{1,n-3},  \label{f18} \\
\eta ^{a_{n-2}} &=&\left( C_{i_{1}\ldots i_{n-1}}^{a}\right) ,  \label{f19}
\end{eqnarray}
together with their conjugated antighosts
\begin{eqnarray}
\mathcal{P}_{a_{0}} &=&\left( \mathcal{P}_{a}^{(1)},\mathcal{P}_{a},\mathcal{%
P}_{ij}^{(1)a},\mathcal{P}_{ij}^{a},P_{a}^{(1)i},P_{a}^{i}\right) ,
\label{f20} \\
\mathcal{P}_{a_{k}} &=&\left( P_{a}^{i_{1}\ldots i_{k+1}},\mathcal{P}%
_{i_{1}\ldots i_{k+2}}^{a}\right) ,\quad k=\overline{1,n-3},  \label{f21} \\
\mathcal{P}_{a_{n-2}} &=&\left( P_{a}^{i_{1}i_{2}\ldots i_{n-1}}\right) .
\label{f22}
\end{eqnarray}
The first set of ghosts respectively corresponds to the first-class
constraints (\ref{f3})--(\ref{f5}), while the other two are due to the
reducibility of the first-class constraint functions. The fields $\eta
^{a_{0}}$ in (\ref{f17}) are fermionic, the fields $\eta ^{a_{k}}$ in (\ref
{f18}) possess the Grassmann parity $\left( k+1\right) \;\mathrm{mod}\mathrm{%
\;}2$, while those in (\ref{f19}) have the Grassmann parity $\left(
n-1\right) \;\mathrm{mod}\mathrm{\;}2$. The ghost number and Grassmann
parity of the antighosts follow from the general rules of the standard
hamiltonian BRST formalism. The ghost number is defined in usual manner as
the difference between the pure ghost number ($\mathrm{pgh}$) and the
antighost number ($\mathrm{agh}$), where
\begin{eqnarray}
\mathrm{pgh}\left( z^{A}\right) &=&0,\;\mathrm{pgh}\left( \eta
^{a_{0}}\right) =1,\quad \mathrm{pgh}\left( \mathcal{P}_{a_{0}}\right) =0,
\label{f23} \\
\mathrm{pgh}\left( \eta ^{a_{k}}\right) &=&k+1,\quad \mathrm{pgh}\left(
\mathcal{P}_{a_{k}}\right) =0,\quad k=\overline{1,n-3},  \label{f24} \\
\mathrm{pgh}\left( \eta ^{a_{n-2}}\right) &=&n-1,\quad \mathrm{pgh}\left(
\mathcal{P}_{a_{n-2}}\right) =0,  \label{f25} \\
\mathrm{agh}\left( z^{A}\right) &=&0,\quad \mathrm{agh}\left( \eta
^{a_{0}}\right) =0,\quad \mathrm{agh}\left( \mathcal{P}_{a_{0}}\right) =1,
\label{f26} \\
\mathrm{agh}\left( \eta ^{a_{k}}\right) &=&0,\quad \mathrm{agh}\left(
\mathcal{P}_{a_{k}}\right) =k+1,\quad k=\overline{1,n-3},  \label{f27} \\
\mathrm{agh}\left( \eta ^{a_{n-2}}\right) &=&0,\quad \mathrm{agh}\left(
\mathcal{P}_{a_{n-2}}\right) =n-1.  \label{f28}
\end{eqnarray}

The BRST charge of this free model takes the form
\begin{eqnarray}
&&\Omega _{0}=\int d^{n-1}x\left( \eta ^{(1)a}G_{a}^{(1)}+\eta
^{a}G_{a}^{(2)}+\eta _{a}^{(1)ij}G_{ij}^{(1)a}+\eta
_{a}^{ij}G_{ij}^{(2)a}+C_{i}^{(1)a}\gamma _{a}^{(1)i}\right.  \nonumber \\
&&\left. +C_{i}^{a}\gamma _{a}^{(2)i}+\sum\limits_{k=3}^{n-1}\left( -\right)
^{k-1}\eta _{a}^{i_{1}i_{2}\ldots i_{k}}\partial _{[i_{1}}\mathcal{P}%
_{i_{2}\ldots i_{k}]}^{a}+\sum\limits_{k=2}^{n-1}\left( -\right)
^{k-1}C_{i_{1}i_{2}\ldots i_{k}}^{a}\partial ^{[i_{1}}P_{a}^{i_{2}\ldots
i_{k}]}\right) ,  \label{f29}
\end{eqnarray}
while the corresponding BRST-invariant Hamiltonian is expressed like
\begin{equation}
H_{0\mathrm{B}}=H_{0}+\int d^{n-1}x\left( \eta ^{(1)a}\mathcal{P}_{a}+\eta
_{a}^{(1)ij}\mathcal{P}_{ij}^{a}+C_{i}^{(1)a}P_{a}^{i}\right) .  \label{f30}
\end{equation}
In general, any function $F$ with $\mathrm{gh}\left( F\right) =0$
that is BRST-closed, $\left[ F,\Omega _{0}\right] =0$, is called
BRST observables. Due to the topological behavior of this model
(the number of physical degrees of freedom is equal to zero), all
the BRST observables are trivial (BRST-exact), meaning that each
of them can be written like $F=\left[
M_{0},\Omega _{0}\right] $, for some fermionic $M_{0}$ with $\mathrm{gh}%
\left( M_{0}\right) =-1$. In particular, the BRST-invariant Hamiltonian is
BRST-exact
\begin{equation}
H_{0\mathrm{B}}=\left[ K_{0},\Omega _{0}\right] ,  \label{f31}
\end{equation}
where, in this situation,
\begin{equation}
K_{0}=\int d^{n-1}x\left( H_{i}^{a}P_{a}^{i}-\frac{1}{2}B_{a}^{ij}\mathcal{P}%
_{ij}^{a}-A_{0}^{a}\mathcal{P}_{a}\right) .  \label{f32}
\end{equation}

The BRST symmetry of the free theory, $s\cdot =\left[ \cdot ,\Omega
_{0}\right] $, splits as
\begin{equation}
s=\delta +\gamma ,  \label{f33}
\end{equation}
where $\delta $ denotes the Koszul-Tate differential ($\mathrm{agh}\left(
\delta \right) =-1$, $\mathrm{pgh}\left( \delta \right) =0$), and $\gamma $
represents the exterior longitudinal derivative ($\mathrm{agh}\left( \gamma
\right) =0$, $\mathrm{pgh}\left( \gamma \right) =1$). These two operators
act on the variables from BRST complex like
\begin{equation}
\delta z^{A}=0,\quad \delta \eta ^{a_{k}}=0,\quad k=\overline{0,n-2},
\label{f34}
\end{equation}
\begin{equation}
\delta \mathcal{P}_{a}^{(1)}=-\pi _{a}^{0},\quad \delta \mathcal{P}%
_{a}=\partial _{i}B_{a}^{0i},\quad \delta P_{a}^{(1)i}=p_{a}^{i},\quad
\delta P_{a}^{i}=-\partial ^{i}\varphi _{a},  \label{f35}
\end{equation}
\begin{equation}
\delta \mathcal{P}_{ij}^{(1)a}=-2\pi _{ij}^{a},\quad \delta \mathcal{P}%
_{ij}^{a}=\partial _{[i}A_{j]}^{a},  \label{f36}
\end{equation}
\begin{equation}
\delta P_{a}^{i_{1}i_{2}\ldots i_{k}}=\left( -\right) ^{k}\partial
^{[i_{1}}P_{a}^{i_{2}\ldots i_{k}]},\quad k=\overline{2,n-1},  \label{f37}
\end{equation}
\begin{equation}
\delta \mathcal{P}_{i_{1}i_{2}\ldots i_{k}}^{a}=\left( -\right) ^{k}\partial
_{[i_{1}}\mathcal{P}_{i_{2}\ldots i_{k}]}^{a},\quad k=\overline{3,n-1},
\label{f38}
\end{equation}
\begin{equation}
\gamma A_{i}^{a}=\partial _{i}\eta ^{a},\quad \gamma A_{0}^{a}=\eta
^{(1)a},\quad \gamma \varphi _{a}=0,\quad \gamma \pi _{a}^{0}=0,\quad \gamma
p_{a}^{i}=0,\quad \gamma \pi _{ij}^{a}=0,  \label{f39}
\end{equation}
\begin{equation}
\gamma B_{a}^{0i}=2\partial _{j}\eta _{a}^{ij},\quad \gamma B_{a}^{ij}=2\eta
_{a}^{(1)ij},\quad \gamma H_{i}^{a}=-C_{i}^{(1)a},\quad \gamma
H_{0}^{a}=\partial ^{i}C_{i}^{a},  \label{f40}
\end{equation}
\begin{equation}
\gamma \eta ^{(1)a}=\gamma \eta ^{a}=\gamma C_{i}^{(1)a}=\gamma \eta
_{a}^{(1)ij}=0,  \label{f41}
\end{equation}
\begin{equation}
\gamma \eta _{a}^{ij}=3\partial _{k}\eta _{a}^{ijk},\quad \gamma
C_{i}^{a}=2\partial ^{j}C_{ij}^{a},  \label{f42}
\end{equation}
\begin{equation}
\gamma \eta _{a}^{i_{1}\ldots i_{k}}=\left( k+1\right) \partial _{i}\eta
_{a}^{ii_{1}\ldots i_{k}},\quad k=\overline{3,n-2},  \label{f43}
\end{equation}
\begin{equation}
\gamma C_{i_{1}\ldots i_{k}}^{a}=-\left( k+1\right) \partial
^{i}C_{ii_{1}\ldots i_{k}}^{a},\quad k=\overline{2,n-2},  \label{f44}
\end{equation}
\begin{equation}
\gamma \eta _{a}^{i_{1}\ldots i_{n-1}}=0,\quad \gamma C_{i_{1}\ldots
i_{n-1}}^{a}=0,  \label{f45}
\end{equation}
\begin{equation}
\gamma \mathcal{P}_{a_{k}}=0,\quad k=\overline{0,n-2}.  \label{f46}
\end{equation}
The last formulas will be employed in the next section at the deformation of
the free theory.

\section{Deformation of the BRST charge\label{defch}}

In this section we solve the equations (\ref{d4})--(\ref{d5}),
etc., that govern the deformation of the BRST charge in the case
of the topological free model under study by relying on
cohomological techniques. As a result, we find that only the
first-order deformation is nontrivial, while its consistency is
equivalent to the existence of a Poisson `two-tensor' (in the
collection indices) depending on the undifferentiated scalar
fields, that must be antisymmetric and fulfills a certain
identity. This two-tensor, together with its derivatives with
respect to the scalar fields, parametrizes the final form of the
BRST charge. Two main types of deformations of the BRST charge are
considered: one that breaks the PT invariance and the other that
preserves it. Although unrelated at the level of the first-order
deformation, these two kinds of solutions become connected when
passing to the higher-order deformations. More precisely, the part
that breaks the PT invariance is initially parametrized by some
completely antisymmetric `tensor' of rank $n$, where $n$ is the
spacetime dimension, which involves only the undifferentiated
scalar fields from the collection. However, the consistency of the
first-order deformation requires that this antisymmetric `tensor'
is expressed precisely via the derivatives of the Poisson
`two-tensor' that parametrizes the PT-invariant solution.

\subsection{First-order deformation\label{firordch}}

Initially, we solve the equation (\ref{d4}), which is responsible for the
first-order deformation of the BRST charge. It takes the local form
\begin{equation}
s\hat{\omega}_{1}=\partial _{i}\hat{j}^{i},  \label{c1}
\end{equation}
for some local $\hat{j}^{i}$. In order to simplify the exposition, we
represent $\hat{\omega}_{1}$ like
\begin{equation}
\hat{\omega}_{1}=\omega _{1}+\bar{\omega}_{1},  \label{xc1}
\end{equation}
where $\omega _{1}$ is the component of the first-order deformation of the
BRST charge that preserves the PT invariance and $\bar{\omega}_{1}$ is the
piece that breaks the PT invariance. The concrete form of $\omega _{1}$ has
been obtained in \cite{noi} and is briefly exposed in the Appendix \ref
{ptinv}. As it has been shown in \cite{noi}, $\omega _{1}$ satisfies
individually an equation of the type (\ref{c1}), and therefore the
decomposition (\ref{xc1}) and the equation (\ref{c1}) require that $\bar{%
\omega}_{1}$ must separately verify a similar equation, i.e.
\begin{equation}
s\bar{\omega}_{1}=\partial _{i}\bar{j}^{i}.  \label{yc1}
\end{equation}
In order to investigate the solutions to this equation, we develop $\bar{%
\omega}_{1}$ according to the antighost number and suppose that the
development stops at a finite order
\begin{equation}
\bar{\omega}_{1}=\stackrel{(0)}{\bar{\omega}}_{1}+\stackrel{(1)}{\bar{\omega}%
}_{1}+\cdots +\stackrel{(J)}{\bar{\omega}}_{1},\quad \mathrm{agh}\left(
\stackrel{(I)}{\bar{\omega}}_{1}\right) =I,\quad \mathrm{gh}\left( \stackrel{%
(I)}{\bar{\omega}}_{1}\right) =1,  \label{c2}
\end{equation}
where the last term can be assumed to be annihilated by $\gamma $
\begin{equation}
\gamma \stackrel{(J)}{\bar{\omega}}_{1}=0.  \label{gama}
\end{equation}
Both results can be shown by adapting the standard lagrangian
arguments from \cite{gen2} to the hamiltonian formulation. Thus,
we need to compute the cohomology of the exterior longitudinal
derivative, $H\left( \gamma \right) $, in order to determine the
piece of highest antighost number in (\ref{c2}).

With the help of the definitions (\ref{f39})--(\ref{f46}) of $\gamma $
acting on the BRST generators, we remark that every local representative of $%
H\left( \gamma \right) $ is generated by
\begin{equation}
\Phi ^{\alpha }=\left( F_{ij}^{a}=\partial _{[i}A_{j]}^{a},\varphi _{a},\pi
_{a}^{0},p_{a}^{i},\pi _{ij}^{a},\partial _{i}B_{a}^{0i}\right) ,  \label{c3}
\end{equation}
(together with their spatial derivatives up to a finite order), by the
antighosts (\ref{f20})--(\ref{f22}) and their spatial derivatives up to a
finite order, as well by the undifferentiated ghosts $\eta ^{a}$, $\eta
_{a}^{i_{1}\ldots i_{n-1}}$, and $C_{i_{1}\ldots i_{n-1}}^{a}$. (The ghosts $%
\eta ^{(1)a}$, $C_{i}^{(1)a}$, and $\eta _{a}^{(1)ij}$, although $\gamma $%
-invariant, are also $\gamma $-exact, and hence trivial in $H\left( \gamma
\right) $. The same is true with respect to the spatial part of the
spacetime derivatives of $\eta ^{a}$, $\eta _{a}^{i_{1}\ldots i_{n-1}}$, and
$C_{i_{1}\ldots i_{n-1}}^{a}$.) In this way, the general, local solution to
the equation (\ref{gama}) can be written (up to trivial, $\gamma $-exact
contributions) as
\begin{equation}
\stackrel{(J)}{\bar{\omega}}_{1}=a_{J}\left( \left[ \Phi ^{\alpha }\right]
,\left[ \mathcal{P}_{a_{k}}\right] _{k=\overline{0,n-2}}\right)
e^{J+1}\left( \eta ^{a},\eta _{a}^{i_{1}\ldots i_{n-1}},C_{i_{1}\ldots
i_{n-1}}^{a}\right) ,  \label{c5}
\end{equation}
where $e^{J+1}\left( \eta ^{a},\eta _{a}^{i_{1}\ldots
i_{n-1}},C_{i_{1}\ldots i_{n-1}}^{a}\right) $ stand for the elements with
pure ghost number equal to $\left( J+1\right) $ of a basis in the space of
the polynomials in the corresponding ghosts, and $a_{J}$ are $\gamma $%
-closed elements of pure ghost number zero, with bounded antighost number, $%
\mathrm{agh}\left( a_{J}\right) =J$. The objects $a_{J}$ play here the role
of `invariant polynomials' \cite{gen1} from the lagrangian approach. The
notation $f\left( \left[ q\right] \right) $ signifies that $f$ depends on $q$
and its spatial derivatives up to a finite order.

The equation (\ref{yc1}) projected on antighost number $\left( J-1\right) $
becomes
\begin{equation}
\delta \stackrel{(J)}{\bar{\omega}}_{1}+\gamma \stackrel{(J-1)}{\bar{\omega}}%
_{1}=\partial _{i}\stackrel{(J)}{\bar{m}}^{i}.  \label{c6}
\end{equation}
Introducing (\ref{c5}) in (\ref{c6}), it follows that a necessary condition
for the existence of (nontrivial) $\stackrel{(J-1)}{\bar{\omega}}_{1}$ is
that the `invariant polynomials' $a_{J}$ from (\ref{c5}) are (nontrivial)
elements of $H_{J}\left( \delta |\tilde{d}\right) $, where the last notation
means the cohomological space of the Koszul-Tate differential modulo the
spatial part of the exterior spacetime derivative in pure ghost number zero
and in strictly positive antighost number $J$%
\begin{equation}
\delta a_{J}=\partial ^{i}n_{i},\quad \mathrm{agh}\left( n_{i}\right)
=J-1,\quad \mathrm{pgh}\left( n_{i}\right) =0.  \label{c6aa}
\end{equation}
Translating the lagrangian results from \cite{gen1} regarding the triviality
of the characteristic cohomology for linear gauge theories at the
hamiltonian level, since our model is $\left( n-2\right) $-order reducible
and the constraint functions are linear in the reduced phase-space
variables, we can state that
\begin{equation}
H_{K}\left( \delta |\tilde{d}\right) =0\quad \mathrm{for\;all}\quad K>n-1.
\label{c7}
\end{equation}
The natural question raises, namely, if the result (\ref{c7}) is still valid
in the space of `invariant polynomials' $H^{\mathrm{inv}}\left( \delta |%
\tilde{d}\right) $, where an element of $H_{J}^{\mathrm{inv}}\left( \delta |%
\tilde{d}\right) $ is defined like in (\ref{c6aa}), but with both
$a_{J}$ and $n_{i}$ `invariant polynomials'. By analyzing what
happens in most gauge theories at the lagrangian level, it is
quite reasonable to assume the validity of (\ref{c7}) in the space
of `invariant polynomials'
\begin{equation}
H_{K}^{\mathrm{inv}}\left( \delta |\tilde{d}\right) =0\quad \mathrm{for\;all}%
\quad K>n-1.  \label{c7a}
\end{equation}
Moreover, (\ref{c7a}) is a consequence of the more general result that if $%
a_{K}$ is an `invariant polynomial' with $\mathrm{agh}\left( a_{K}\right)
=K\geq n-1$, which is trivial in $H_{K}\left( \delta |\tilde{d}\right) $, $%
a_{K}=\delta b_{K}+\partial ^{i}m_{i}$, with $\mathrm{agh}\left(
b_{K}\right) =K+1$ and $\mathrm{agh}\left( m_{i}\right) =K$, then it can be
taken to be trivial also in $H_{K}^{\mathrm{inv}}\left( \delta |\tilde{d}%
\right) $, i.e., both $b_{K}$ and $m_{i}$ can be taken to be `invariant
polynomials'.

The previous results on $H\left( \delta |\tilde{d}\right) $ and $H^{\mathrm{%
inv}}\left( \delta |\tilde{d}\right) $ in strictly positive antighost
numbers are important because they control the obstructions to removing the
antighosts from the first-order deformation of the BRST charge. More
precisely, one can successively eliminate all the pieces of antighost number
strictly greater than $\left( n-1\right) $ from $\bar{\omega}_{1}$ by adding
only trivial terms, so one can take, without loss of nontrivial objects, the
condition $J\leq n-1$ in the decomposition (\ref{c2}). Moreover, the last
representative can always be taken to belong to $H\left( \gamma \right) $,
with the corresponding `invariant polynomial' a nontrivial object from $%
H_{J}^{\mathrm{inv}}\left( \delta |\tilde{d}\right) $ for $J>1$
and respectively from $H_{1}\left( \delta |\tilde{d}\right) $ if
$J=1$.

Consequently, we can assume that $J=n-1$ in (\ref{c2})
\begin{equation}
\bar{\omega}_{1}=\stackrel{(0)}{\bar{\omega}}_{1}+\stackrel{(1)}{\bar{\omega}%
}_{1}+\cdots +\stackrel{(n-1)}{\bar{\omega}}_{1},  \label{c8}
\end{equation}
with $\stackrel{(n-1)}{\bar{\omega}}_{1}$ given by (\ref{c5}) for $J=n-1$
and $a_{n-1}$ a nontrivial element from $H_{n-1}^{\mathrm{inv}}\left( \delta
|\tilde{d}\right) $. After some computation, we find that the most general
representative of both $H_{n-1}\left( \delta |\tilde{d}\right) $ and $%
H_{n-1}^{\mathrm{inv}}\left( \delta |\tilde{d}\right) $ can be expressed
like
\begin{eqnarray}
&&a_{n-1}^{i_{1}\ldots i_{n-1}}=\frac{\partial U}{\partial \varphi _{a}}%
P_{a}^{i_{1}\ldots i_{n-1}}+\sum\limits_{p=2}^{n-1}\sum\limits_{1\leq
j_{1}\leq j_{2}\leq \cdots \leq j_{p}<n-1}\frac{\partial ^{p}U}{\partial
\varphi _{a_{1}}\partial \varphi _{a_{2}}\cdots \partial \varphi _{a_{p}}}%
\times  \nonumber \\
&&\times P_{a_{1}}^{[i_{1}\ldots i_{j_{1}}}P_{a_{2}}^{i_{j_{1}+1}\ldots
i_{j_{1}+j_{2}}}\cdots P_{a_{p-1}}^{i_{j_{1}+\cdots +j_{p-2}+1}\ldots
i_{j_{1}+\cdots +j_{p-1}}}P_{a_{p}}^{i_{j_{1}+\cdots +j_{p-1}+1}\ldots
i_{n-1}]},  \label{c9}
\end{eqnarray}
where $U$ is an arbitrary function involving only the undifferentiated
scalar fields $\varphi _{a}$, and $j_{p}$ means
\begin{equation}
j_{p}=n-1-\left( j_{1}+j_{2}+\cdots +j_{p-1}\right) .  \label{notlong}
\end{equation}
Taking into account the definitions (\ref{f34})--(\ref{f38}) of the
Koszul-Tate differential, one can prove the recursive relations
\begin{equation}
\delta a_{k}^{i_{1}i_{2}\ldots i_{k}}=\left( -\right) ^{k}\partial
^{[i_{1}}a_{k-1}^{i_{2}\ldots i_{k}]},\quad k=\overline{1,n-1},
\label{c10}
\end{equation}
where for $k=\overline{2,n-2}$ we have
\begin{eqnarray}
&&a_{k}^{i_{1}\ldots i_{k}}=\frac{\partial U}{\partial \varphi _{a}}%
P_{a}^{i_{1}\ldots i_{k}}+\sum\limits_{q=2}^{k}\sum\limits_{1\leq j_{1}\leq
j_{2}\leq \cdots \leq j_{q}<k}\frac{\partial ^{q}U}{\partial \varphi
_{a_{1}}\partial \varphi _{a_{2}}\cdots \partial \varphi _{a_{q}}}\times
\nonumber \\
&&\times P_{a_{1}}^{[i_{1}\ldots i_{j_{1}}}P_{a_{2}}^{i_{j_{1}+1}\ldots
i_{j_{1}+j_{2}}}\cdots P_{a_{q-1}}^{i_{j_{1}+\cdots +j_{q-2}+1}\ldots
i_{j_{1}+\cdots +j_{q-1}}}P_{a_{q}}^{i_{j_{1}+\cdots +j_{q-1}+1}\ldots
i_{k}]},  \label{c11}
\end{eqnarray}
while for $k=1$ and respectively $k=0$ we obtain
\begin{equation}
a_{1}^{i}=\frac{\partial U}{\partial \varphi _{a}}P_{a}^{i},\quad a_{0}=U.
\label{c12}
\end{equation}
In (\ref{c11}) we used the notation $j_{q}=k-\left( j_{1}+j_{2}+\cdots
+j_{q-1}\right) $. Now, we can completely determine the last component in (%
\ref{c8}). The elements of pure ghost number equal to $n$, $e^{n}\left( \eta
^{a},\eta _{a}^{i_{1}\ldots i_{n-1}},C_{i_{1}\ldots i_{n-1}}^{a}\right) $,
are given by
\begin{equation}
e^{n}:\left( \eta ^{a}C_{i_{1}\ldots i_{n-1}}^{b},\eta ^{a}\eta ^{b}\eta
_{c}^{i_{1}\ldots i_{n-1}},\eta ^{a_{1}}\eta ^{a_{2}}\cdots \eta
^{a_{n}}\right)  \label{c13}
\end{equation}
for all $n\geq 4$\footnote{For $n=4$ there is an extra possibility
because $\eta _{a}^{i_{1}\cdots i_{n-1}}\rightarrow \eta
_{a}^{ijk}$, with $pgh\left( \eta _{a}^{ijk}\right) =2$, and so we
have a supplementary element of the basis in the ghosts at pure
ghost number $n=4$, namely, $\eta _{a}^{ijk}\eta _{b}^{i^{\prime
}j^{\prime }k^{\prime }}$. However, this element can be discarded
\cite {ijmpa}, so finally (\ref{c13}) still covers all the
investigated situations.}. It means that the piece of highest
antighost number in the first-order deformation is fully
determined once we `glue' (\ref{c9}) to (\ref{c13}) like in
(\ref{c5}). The last component of $e^{n}$ needs the adjustment of
a completely antisymmetric constant $K_{i_{1}\ldots i_{n-1}}$ in
order to match (\ref{c9}), which can only be, by `covariance'
arguments, proportional to the spatial part of the completely
antisymmetric symbol in $n$ dimensions, $\varepsilon
_{0i_{1}\ldots i_{n-1}}$. Even if we `force' the introduction of
additional antisymmetric symbols in the components of
$\hat{\omega}_{1}$ involving the first two elements in
(\ref{c13}), we finally obtain that such terms are always
proportional with some objects that contain no antisymmetric
symbols, like in (\ref{pt14}). In conclusion, there is no
possibility to construct pieces from
$\stackrel{(n-1)}{\bar{\omega}}_{1}$ that involve either of the
first two elements in (\ref{c13}). Such terms can only bring
contributions to the element $\stackrel{(n-1)}{\omega }_{1}$ of
highest antighost number in the first-order deformation of the
BRST charge that preserves the PT invariance, $\omega _{1}$.

As we have stated in the beginning of this section, here we focus only on
the interactions that break the PT invariance and which, by virtue of the
above discussion, can be generated just by the third element in (\ref{c13}),
such that we can write
\begin{equation}
\stackrel{(n-1)}{\bar{\omega}}_{1}=\frac{\left( -\right) ^{\left[ \frac{n}{2}%
\right] }}{n!}\epsilon _{0i_{1}\ldots i_{n-1}}N_{a_{1}\ldots
a_{n}}^{i_{1}\ldots i_{n-1}}\eta ^{a_{1}}\cdots \eta ^{a_{n}},  \label{fpt2}
\end{equation}
where $\left[ \frac{n}{2}\right] $ denotes the integer part of $\frac{n}{2}$%
. The element $N_{a_{1}\ldots a_{n}}^{i_{1}\ldots i_{n-1}}$ in
(\ref{fpt2}) results from $a_{n-1}^{i_{1}i_{2}\ldots i_{n-1}}$ in
(\ref{c9}) where we replace the function $U$ depending only on the
undifferentiated scalar fields with a completely antisymmetric
`tensor' of rank $n$, $N_{a_{1}\ldots a_{n}}$, also involving just
the $\varphi _{a}$'s
\begin{eqnarray}
&&N_{a_{1}\ldots a_{n}}^{i_{1}\ldots i_{n-1}}=\frac{\partial N_{a_{1}\ldots
a_{n}}}{\partial \varphi _{b}}P_{b}^{i_{1}i_{2}\ldots
i_{n-1}}+\sum\limits_{p=2}^{n-1}\sum\limits_{1\leq j_{1}\leq j_{2}\leq
\cdots \leq j_{p}<n-1}\frac{\partial ^{p}N_{a_{1}\ldots a_{n}}}{\partial
\varphi _{b_{1}}\partial \varphi _{b_{2}}\cdots \partial \varphi _{b_{p}}}%
\times  \nonumber \\
&&\times P_{b_{1}}^{[i_{1}\ldots i_{j_{1}}}P_{b_{2}}^{i_{j_{1}+1}\ldots
i_{j_{1}+j_{2}}}\cdots P_{b_{p-1}}^{i_{j_{1}+\cdots +.j_{p-2}+1}\ldots
i_{j_{1}+\cdots +j_{p-1}}}P_{b_{p}}^{i_{j_{1}+\cdots +j_{p-1}+1}\ldots
i_{n-1}]}.  \label{fpt3}
\end{eqnarray}
The supplementary numerical factor from (\ref{fpt2}) has been added for
further convenience.

Introducing the relation (\ref{fpt2}) into the equation (\ref{c6}) for $J=n-1
$ and using the definitions (\ref{f34})--(\ref{f46}), we infer that the
piece with the antighost number equal to $\left( n-2\right) $ from the
first-order deformation of the BRST charge that breaks the PT invariance
reads as
\begin{equation}
\stackrel{(n-2)}{\bar{\omega}}_{1}=\frac{\left( -\right) ^{\left[ \frac{n}{2}%
\right] +1}}{\left( n-1\right) !}\sum\limits_{p\geq 1}\left( -\right)
^{p}\epsilon _{0i_{1}\ldots i_{n-1-p}\ldots i_{n-1}}N_{a_{1}\ldots
a_{n}}^{[i_{1}\ldots i_{n-1-p}}\mathcal{P}^{a_{1}i_{n-p}\ldots i_{n-1}]}\eta
^{a_{2}}\cdots \eta ^{a_{n}}.  \label{fpt4}
\end{equation}
If we take into account the decomposition (\ref{f33}) of the free BRST
differential and insert the expansion (\ref{c8}) into the equation (\ref{yc1}%
), it follows that the component $\stackrel{(n-3)}{\bar{\omega}}_{1}$ is
solution to the equation (\ref{c6}) with $J\rightarrow n-2$. Substituting
the solution (\ref{fpt4}) into this equation and using the definitions (\ref
{f34})--(\ref{f46}), after some computation we find that
\begin{eqnarray}
\stackrel{(n-3)}{\bar{\omega}}_{1} &=&\frac{\left( -\right) ^{\left[ \frac{n%
}{2}\right] +2}}{\left( n-2\right) !}\sum\limits_{p_{1}\geq p_{2}\geq
1}\left( -\right) ^{p_{1}}\epsilon _{0i_{1}\ldots i_{n-1-\left(
p_{1}+p_{2}\right) }\ldots i_{n-1}}N_{a_{1}\ldots a_{n}}^{[i_{1}\ldots
i_{n-1-\left( p_{1}+p_{2}\right) }}\times   \label{fpt5} \\
&&\times \mathcal{P}^{a_{1}i_{n-\left( p_{1}+p_{2}\right) }\ldots
i_{n-1-p_{2}}}\mathcal{P}^{a_{2}i_{n-p_{2}}\ldots i_{n-1}]}\eta
^{a_{3}}\cdots \eta ^{a_{n}}.  \nonumber
\end{eqnarray}
In a similar manner we solve the equations that govern the terms of
antighost number $\left( n-m\right) $ from $\bar{\omega}_{1}$ with $m=%
\overline{4,n-2}$, which are expressed by (\ref{c6}) with $J\rightarrow n-m+1
$, and get that
\begin{eqnarray}
\stackrel{(n-m)}{\bar{\omega}}_{1} &=&\frac{\left( -\right) ^{\left[ \frac{n%
}{2}\right] +m-1}}{\left( n-m+1\right) !}\sum\limits_{p_{1}\geq p_{2}\geq
\cdots \geq p_{m-1}\geq 1}\left( -\right) ^{\Xi }\epsilon _{0i_{1}\ldots
i_{n-1-\left( p_{1}+p_{2}+\cdots +p_{m-1}\right) }\ldots i_{n-1}}\times
\nonumber \\
&&\times N_{a_{1}\ldots a_{n}}^{[i_{1}\ldots i_{n-1-\left( p_{1}+\cdots
+p_{m-1}\right) }}\mathcal{P}^{a_{1}i_{n-\left( p_{1}+p_{2}+\cdots
+p_{m-1}\right) }\ldots i_{n-1-\left( p_{2}+\cdots +p_{m-1}\right) }}\times
\nonumber \\
&&\times \mathcal{P}^{a_{2}i_{n-\left( p_{2}+\cdots +p_{m-1}\right) }\ldots
i_{n-1-\left( p_{3}+\cdots +p_{m-1}\right) }}\cdots \mathcal{P}%
^{a_{m-1}i_{n-p_{m-1}}\ldots i_{n-1}]}\eta ^{a_{m}}\cdots \eta ^{a_{n}},
\label{fpt6}
\end{eqnarray}
where we made the notations $M=\left( \left[ \frac{m}{2}\right] -1\right) $
and $\Xi \equiv p_{1}+p_{3}+\cdots +p_{2M+1}$. In the equations (\ref{fpt4}%
)--(\ref{fpt6}) we denoted by $\mathcal{P}^{ai}$ the spatial part of the
collection of one-forms $A^{a\mu }$%
\begin{equation}
\mathcal{P}^{ai}\equiv A^{ai}  \label{notsup}
\end{equation}
in order to write the pieces that compose the first-order deformation of the
BRST charge $\bar{\omega}_{1}$ in a brief, compact manner. We proceed in the
same way with the equation (\ref{c6}) for $J\rightarrow 2$ and $J\rightarrow
1$, and determine the elements of antighost number one and respectively zero
from $\bar{\omega}_{1}$ in the form
\begin{eqnarray}
\stackrel{(1)}{\bar{\omega}}_{1} &=&\frac{1}{2}\epsilon _{0i_{1}\ldots
i_{n-1}}\left( -N_{a_{1}\ldots
a_{n}}^{[i_{1}}A^{a_{1}i_{2}}A^{a_{2}i_{3}}\cdots A^{a_{n-2}i_{n-1}]}\right.
\nonumber \\
&&\left. +N_{a_{1}\ldots a_{n}}\mathcal{P}%
^{a_{1}[i_{1}i_{2}}A^{a_{2}i_{3}}A^{a_{3}i_{4}}\cdots
A^{a_{n-2}i_{n-1}]}\right) \eta ^{a_{n-1}}\eta ^{a_{n}},  \label{fpt7}
\end{eqnarray}
\begin{equation}
\stackrel{(0)}{\bar{\omega}}_{1}=\epsilon _{0i_{1}\ldots
i_{n-1}}N_{aa_{1}\ldots a_{n-1}}A^{a_{1}i_{1}}\cdots A^{a_{n-1}i_{n-1}}\eta
^{a},  \label{fpt8}
\end{equation}
where we reverted to the one-form notation.

In consequence, so far we have computed the first-order deformation of the
BRST charge that breaks the PT invariance like
\begin{equation}
\bar{\omega}_{1}=\sum\limits_{k=0}^{n-1}\stackrel{(k)}{\bar{\omega}}_{1},
\label{fpt9}
\end{equation}
such that the full first-order deformation of the BRST charge is
\begin{equation}
\hat{\Omega}_{1}=\int d^{n-1}x\left( \omega _{1}+\bar{\omega}_{1}\right) ,
\label{ww1}
\end{equation}
with $\omega _{1}$ given in the Appendix \ref{ptinv}. We emphasize
that the solutions $\stackrel{(k)}{\bar{\omega}}_{1}$ obtained in
the above also include, for all $k<n-1$, the solutions
corresponding to the associated `homogeneous' equations $\gamma
\stackrel{(k)}{\bar{\omega}^{\prime }}_{1}=0$. In order to
simplify the exposition we avoided the discussion regarding the
selection procedure of these solutions such as to comply with
obtaining some consistent components of the first-order
deformation of the BRST charge at each value of the antighost
number. It is however interesting to note that this procedure
allows no new functions of the scalar fields beside
$N_{a_{1}\ldots a_{n}}$ to parametrize the solution
$\bar{\omega}_{1}$.

\subsection{Higher-order deformations}

Our next concern is to analyze the existence of higher-order deformations of
the BRST charge. In view of this, we make the notations $\hat{\Omega}%
_{2}=\int d^{n-1}x\,b$ and $\left[ \hat{\Omega}_{1},\hat{\Omega}_{1}\right]
=\int d^{n-1}x\,\Delta $, and observe that the equation (\ref{d5}), which
governs the second-order deformation of the BRST charge, takes the local
form
\begin{equation}
\Delta =-2sb+\partial _{i}m^{i}.  \label{fpt10}
\end{equation}
We mention that at this stage the entire BRST charge (\ref{ww1}) must be
taken into account, and not only the PT-breaking component $\bar{\omega}_{1}$%
. In view of the results from the previous subsection, combined with those
from the Appendix \ref{ptinv}, direct computation finally leads to
\begin{eqnarray}
\Delta &=&K^{abc}t_{abc}+\sum\limits_{k=1}^{n-1}K_{a_{1}a_{2}\ldots
a_{k}}^{abc}\frac{\partial ^{k}t_{abc}}{\partial \varphi _{a_{1}}\partial
\varphi _{a_{2}}\cdots \partial \varphi _{a_{k}}}  \nonumber \\
&&+K_{d}^{abc}t_{abc}^{d}+\sum\limits_{k=1}^{n-1}K_{d,\;a_{1}a_{2}\ldots
a_{k}}^{abc}\frac{\partial ^{k}t_{abc}^{d}}{\partial \varphi
_{a_{1}}\partial \varphi _{a_{2}}\cdots \partial \varphi _{a_{k}}}  \nonumber
\\
&&+K^{a_{1}\ldots a_{n+1}}t_{a_{1}\ldots
a_{n+1}}+\sum\limits_{k=1}^{n-1}K_{b_{1}\ldots b_{k}}^{a_{1}\ldots a_{n+1}}%
\frac{\partial ^{k}t_{a_{1}\ldots a_{n+1}}}{\partial \varphi
_{b_{1}}\partial \varphi _{b_{2}}\cdots \partial \varphi _{b_{k}}},
\label{fpt11}
\end{eqnarray}
where
\begin{eqnarray}
t_{abc} &=&W_{ec}M_{ab}^{c}+W_{ea}\frac{\partial W_{bc}}{\partial \varphi
_{e}}+W_{eb}\frac{\partial W_{ca}}{\partial \varphi _{e}},  \label{uzx} \\
t_{abc}^{d} &=&W_{e[a}\frac{\partial M_{bc]}^{d}}{\partial \varphi _{e}}%
+M_{e[a}^{d}M_{bc]}^{e},  \label{uzy} \\
t_{a_{1}\ldots a_{n+1}} &=&\frac{\partial N_{[a_{1}\ldots a_{n}}}{\partial
\varphi _{b}}W_{a_{n+1}]b}+M_{[a_{1}a_{2}}^{b}N_{a_{3}\ldots a_{n+1}]b}.
\label{fpt12}
\end{eqnarray}
All the objects $K^{abc}$, $K_{d}^{abc}$, $K_{a_{1}a_{2}\ldots a_{k}}^{abc}$%
, $K_{d,\;a_{1}a_{2}\ldots a_{k}}^{abc}$, $K^{a_{1}\ldots a_{n}}$, and $%
K_{b_{1}\ldots b_{k}}^{a_{1}\ldots a_{n}}$ are polynomials that involve only
undifferentiated ghosts, antighosts, and fields $B_{a}^{0i}$ and $A_{i}^{a}$%
, none of them being BRST-exact. On the other hand, the equation (\ref{fpt10}%
) requires that $\Delta $ is $s$-exact modulo $\tilde{d}$, and, in fact,
since $\Delta $ contains no derivatives, it demands that $\Delta $ must be $%
s $-exact. Since neither of the terms in (\ref{fpt11}) is so, it results
that the consistency of the first-order deformation of the BRST charge asks
that $\Delta $ must vanish. This takes place if and only if the equations
\begin{eqnarray}
t_{abc} &=&0,\quad t_{abc}^{d}=0,  \label{fpt14} \\
t_{a_{1}\ldots a_{n+1}} &=&0  \label{ww2}
\end{eqnarray}
are simultaneously satisfied. As it has been shown in \cite{noi}, the
general solution to the equations (\ref{fpt14}) is of the form
\begin{equation}
M_{ab}^{c}=\frac{\partial W_{ab}}{\partial \varphi _{c}},  \label{zxw}
\end{equation}
where $W_{ab}$ is an antisymmetric `two-tensor' in the undifferentiated
scalar fields, subject to the identities
\begin{equation}
W_{e[a}\frac{\partial W_{bc]}}{\partial \varphi _{c}}=0.  \label{zsa}
\end{equation}
Due to the fact that the antisymmetric functions $W_{ab}$ depend only on the
undifferentiated scalar fields from the collection and verify the identities
(\ref{zsa}), they can be interpreted as the components of the Poisson
two-tensor of a Poisson manifold with the target space locally parametrized
by the scalar fields. Under these circumstances, the general solution of (%
\ref{ww2}) can be represented like
\begin{equation}
N_{a_{1}\ldots a_{n}}=f_{b[a_{1}\ldots a_{n-2}}\frac{\partial
W_{a_{n-1}a_{n}]}}{\partial \varphi _{b}},  \label{fpt15}
\end{equation}
where $f_{a_{1}\ldots a_{n}}$ are some completely antisymmetric
constants. Consequently, we can take the second-order deformation
of the BRST charge to vanish, $\hat{\Omega}_{2}=0$, and, in fact,
all the higher-order deformation equations are satisfied with the
choice
\begin{equation}
\hat{\Omega}_{k}=0\quad \mathrm{for\;all}\quad k\geq 2.  \label{fpt16}
\end{equation}
In conclusion, the deformed BRST charge, consistent to all orders in the
coupling constant, simply reduces to the sum between the free BRST charge
and the first-order deformation
\begin{equation}
\hat{\Omega}=\Omega _{0}+g\hat{\Omega}_{1}=\Omega _{0}+g\int
d^{n-1}x\sum\limits_{k=0}^{n-1}\left( \stackrel{\left( k\right) }{\omega _{1}%
}+\stackrel{\left( k\right) }{\bar{\omega}_{1}}\right) =\Omega _{0}+g\left(
\Omega _{1}+\bar{\Omega}_{1}\right) ,  \label{fpt16'}
\end{equation}
where $\stackrel{\left( k\right) }{\bar{\omega}_{1}}$ and $\stackrel{\left(
k\right) }{\omega _{1}}$ are listed in (\ref{fpt2}), (\ref{fpt4})--(\ref
{fpt6}), (\ref{fpt7})--(\ref{fpt8}) and respectively in (\ref{pt14}), (\ref
{pt20})--(\ref{pt24}), with the observation that $M_{ab}^{c}$ and $%
N_{a_{1}\ldots a_{n}}$ must be replaced with (\ref{zxw}) and (\ref{fpt15}),
while $W_{ab}$ are assumed to verify the identities (\ref{zsa}).

\section{Deformation of the BRST-invariant Hamiltonian}

We now turn our attention to the BRST-invariant Hamiltonian
(\ref{f30}), whose deformation is stipulated by the equations
(\ref{d9})--(\ref{d10}), etc. We will prove that the deformed
BRST-invariant Hamiltonian, just like the BRST charge, stops at
order one in the coupling constant and, moreover, is trivial
(BRST-exact) with respect to the fully deformed BRST symmetry,
which confirms the preservation of the topological behavior also
at the level of the interacting theory.

Initially, we approach the equation (\ref{d9}), associated with its
first-order deformation. Inserting (\ref{f31}) in (\ref{d9}) and using (\ref
{d4}) and the Jacobi identity with respect to the Dirac bracket, we find
that (\ref{d9}) is in fact equivalent to the equation $\left[ \hat{H}%
_{1}-\left[ K_{0},\hat{\Omega}_{1}\right] ,\Omega _{0}\right] =0$, which
shows that $\hat{H}_{1}-\left[ K_{0},\hat{\Omega}_{1}\right] $ is nothing
but a BRST observable of the free theory. As it has been argued in Section
\ref{free}, all the BRST observables are in this case trivial (BRST-exact),
or, in other words, they belong to the same equivalence class as the trivial
observable zero. In consequence, we can take
\begin{equation}
\hat{H}_{1}=\left[ K_{0},\hat{\Omega}_{1}\right] ,  \label{h1}
\end{equation}
where the function $K_{0}$ is displayed in (\ref{f32}). Next, we split $\hat{%
H}_{1}$ like the first-order deformation of the BRST charge, as the sum
between the PT-invariant component $H_{1}$ and the PT-breaking part $\bar{H}%
_{1}$%
\begin{equation}
\hat{H}_{1}=H_{1}+\bar{H}_{1}.  \label{wq1}
\end{equation}
Recalling the similar decomposition of the first-order deformation of the
BRST charge, we consequently get that
\begin{equation}
H_{1}+\bar{H}_{1}=\left[ K_{0},\Omega _{1}\right] +\left[ K_{0},\bar{\Omega}%
_{1}\right] .  \label{wq2}
\end{equation}
It has been shown in \cite{noi} that
\begin{equation}
H_{1}=\left[ K_{0},\Omega _{1}\right] ,  \label{wq3}
\end{equation}
where the non-integrated density of $H_{1}$ has the expression
\begin{eqnarray}
h_{1}&=&-W_{ab}H_{\mu }^{a}A^{b\mu }-\frac{1}{2}M_{ab}^{c}A_{\mu
}^{a}A_{\nu
}^{b}B_{c}^{\mu \nu }  \nonumber \\
&&-M_{ab}^{c}\left( \frac{1}{2}B_{c}^{ij}\eta ^{a}\mathcal{P}%
_{ij}^{b}+A_{0}^{a}\mathcal{P}_{ij}^{b}\eta _{c}^{ij}+A_{0}^{a}\eta ^{b}%
\mathcal{P}_{c}\right)  \nonumber \\
&&+\frac{\partial W_{ab}}{\partial \varphi _{c}}P_{c}^{i}\left(
H_{i}^{a}\eta ^{b}+C_{i}^{a}A_{0}^{b}\right)  \nonumber \\
&&+\frac{\partial M_{ab}^{c}}{\partial \varphi _{d}}P_{di}\left( \eta
^{a}A_{j}^{b}B_{c}^{ij}-\eta ^{a}A_{0}^{b}B_{c}^{0i}+2A_{0}^{a}A_{j}^{b}\eta
_{c}^{ij}\right)  \nonumber \\
&&+\frac{1}{4}\left( \frac{\partial M_{ab}^{c}}{\partial \varphi _{d}}%
P_{dij}+\frac{\partial ^{2}M_{ab}^{c}}{\partial \varphi _{d}\partial \varphi
_{e}}P_{di}P_{ej}\right) \eta ^{a}\eta ^{b}B_{c}^{ij}  \nonumber \\
&&+\sum\limits_{k=2}^{n-1}A_{0}^{a}\frac{\partial ^{L}\stackrel{(k)}{\omega }%
_{1}}{\partial \eta ^{a}},  \label{h2}
\end{eqnarray}
and the notation $\partial ^{L}/\partial \eta ^{a}$ for the left derivative
with respect to $\eta ^{a}$ was employed. In (\ref{h2}) and further, $%
M_{ab}^{c}$ takes the form (\ref{zxw}). By means of the relations (\ref{wq2}%
)--(\ref{wq3}), one finds that $\bar{H}_{1}$ checks the equation
\begin{equation}
\bar{H}_{1}=\left[ K_{0},\bar{\Omega}_{1}\right] .  \label{wq4}
\end{equation}
On behalf of $\bar{\Omega}_{1}$ given in (\ref{wq4}), we determine
that the non-integrated density of $\bar{H}_{1}$ reads as
\begin{eqnarray}
\bar{h}_{1} &=&\epsilon _{0i_{1}\ldots i_{n-1}}N_{a_{1}\ldots
a_{n}}A^{a_{1}0}A^{a_{2}i_{2}}\cdots A^{a_{n}i_{n-1}}  \nonumber \\
&&-\epsilon _{0i_{1}\ldots i_{n-1}}\eta ^{a_{1}}\left( N_{a_{1}\ldots
a_{n}}A^{a_{2}0}\mathcal{P}^{a_{3}[i_{1}i_{2}}A^{a_{4}i_{3}}\cdots
A^{a_{n}i_{n-1}]}\right.  \nonumber \\
&&\left. -\frac{\partial N_{a_{1}\ldots a_{n}}}{\partial \varphi _{b}}%
A^{a_{2}0}P_{b}^{[i_{1}}A^{a_{3}i_{2}}\cdots A^{a_{n}i_{n-1}]}\right)
+\sum\limits_{k=2}^{n-1}A_{0}^{a}\frac{\partial ^{L}\stackrel{(k)}{\bar{%
\omega}}_{1}}{\partial \eta ^{a}},  \label{fpt18}
\end{eqnarray}
where in (\ref{fpt18}) and in what follows $N_{a_{1}\ldots a_{n}}$ takes the
form (\ref{fpt15}).

In the next step we investigate the second-order deformation of the
BRST-invariant Hamiltonian, subject to the equation (\ref{d10}). We observe
that the third term in the right-hand side of the equation (\ref{d10})
vanishes since $\hat{\Omega}_{2}=0$. Making use of (\ref{h1}) and employing
once more the Jacobi identity with respect to the Dirac bracket, it is easy
to see that the second term in the right-hand side of (\ref{d10}) can be
written like
\begin{equation}
\left[ \hat{H}_{1},\hat{\Omega}_{1}\right] =\frac{1}{2}\left[ K_{0},\left[
\hat{\Omega}_{1},\hat{\Omega}_{1}\right] \right] ,  \label{h3}
\end{equation}
and it vanishes according to the fact (established in the previous section)
that $\left[ \hat{\Omega}_{1},\hat{\Omega}_{1}\right] =0$. Then, we can set
\begin{equation}
\hat{H}_{2}=0,  \label{h5}
\end{equation}
which leads to the fact that the remaining higher-order equations are
satisfied for
\begin{equation}
\hat{H}_{k}=0,\quad k>2.  \label{h4}
\end{equation}
As a consequence, we can write the fully deformed BRST-invariant Hamiltonian
like
\begin{equation}
H_{\mathrm{B}}=H_{0\mathrm{B}}+g\left( H_{1}+\bar{H}_{1}\right) =H_{0\mathrm{%
B}}+g\hat{H}_{1},  \label{h6}
\end{equation}
but also, taking into account (\ref{f31}), (\ref{fpt16'}), and
(\ref{h1})
\begin{equation}
H_{\mathrm{B}}=\left[ K_{0},\hat{\Omega}\right] .  \label{h7}
\end{equation}
The last formula confirms the topological behavior of the
interacting model. It stresses that $H_{\mathrm{B}}$ is not only
invariant with respect to the deformed hamiltonian BRST symmetry,
but also exact. This ends the deformation procedure of the
BRST-invariant Hamiltonian for the free theory under study.

\section{Description of the interacting model\label{int}}

With the deformed BRST charge and BRST-invariant Hamiltonian at hand, in the
sequel we will be able to identify the main ingredients of the hamiltonian
formulation for the resulting interacting model and, on these grounds, also
the associated coupled lagrangian action and its gauge transformations. We
recall that the deformed BRST charge is given in (\ref{fpt16'}), with the
corresponding components listed in (\ref{fpt2}), (\ref{fpt4})--(\ref{fpt6}),
(\ref{fpt7})--(\ref{fpt8}) and respectively in (\ref{pt14}), (\ref{pt20})--(%
\ref{pt24}), while the deformed BRST-invariant Hamiltonian reads
as in (\ref {h6}), with the non-integrated densities of $H_{1}$
and $\bar{H}_{1}$ given
in (\ref{h2}) and (\ref{fpt18}). The functions $M_{ab}^{c}$ and $%
N_{a_{1}\ldots a_{n}}$ must be replaced with (\ref{zxw}) and (\ref{fpt15}),
while $W_{ab}$ are assumed to verify the identities (\ref{zsa}).

It is well known that from the BRST charge and the BRST-invariant
Hamiltonian one can withdraw the entire hamiltonian formulation of a gauge
theory. Thus, from the terms of antighost number zero from (\ref{fpt16'}) we
see that only the secondary first-class constraints of the interacting model
are deformed like
\begin{eqnarray}
\tilde{G}_{a}^{(2)}&\equiv &-\left( D_{i}\right)
_{a}^{\;\;b}B_{b}^{0i}+gW_{ab}H_{0}^{b}+g\epsilon _{0i_{1}\ldots
i_{n-1}}\times  \nonumber \\
&&\times N_{aa_{1}\ldots
a_{n-1}}A^{a_{1}i_{1}}A^{a_{2}i_{2}}\cdots
A^{a_{n-1}i_{n-1}}\approx 0,  \label{fpt26}
\end{eqnarray}
\begin{equation}
\bar{G}_{ij}^{(2)a}\equiv -\bar{F}_{ij}^{a}\approx 0,  \label{fpt27}
\end{equation}
\begin{equation}
\bar{\gamma}_{a}^{(2)i}\equiv D^{i}\varphi _{a}\approx 0,  \label{fpt28}
\end{equation}
while the primary ones are not affected by the deformation procedure, being
given by (\ref{f3})--(\ref{f5}). In the above we used the notations
\begin{eqnarray}
\left( D_{i}\right) _{a}^{\;\;b} &=&\delta _{a}^{b}\partial _{i}+g\frac{%
\partial W_{ac}}{\partial \varphi _{b}}A_{i}^{c},  \label{pt47} \\
\bar{F}_{ij}^{a} &=&\partial _{[i}A_{j]}^{a}+g\frac{\partial W_{bc}}{%
\partial \varphi _{a}}A_{i}^{b}A_{j}^{c},  \label{pt48} \\
D^{i}\varphi _{a} &=&\partial ^{i}\varphi _{a}+gW_{ab}A^{bi}.  \label{pt49}
\end{eqnarray}
The pieces of antighost number one in (\ref{fpt16'}) reveal that only the
Dirac brackets between the secondary first-class constraints are modified as
\begin{eqnarray}
\left[ \tilde{G}_{a}^{(2)},\tilde{G}_{b}^{(2)}\right] &=&-g\left( \frac{%
\partial W_{ab}}{\partial \varphi _{c}}\tilde{G}_{c}^{(2)}-\frac{\partial
^{2}W_{ab}}{\partial \varphi _{c}\partial \varphi _{d}}B_{d0i}\bar{\gamma}%
_{c}^{(2)i}\right)  \nonumber \\
&&+g\epsilon _{0i_{1}\ldots i_{n-1}}\left( \frac{\partial N_{abc_{1}\ldots
c_{n-2}}}{\partial \varphi _{d}}\bar{\gamma}_{d}^{\left( 2\right)
[i_{1}}A^{c_{1}i_{2}}\cdots A^{c_{n-2}i_{n-1}]}\right.  \nonumber \\
&&\left. -N_{abc_{1}\ldots c_{n-2}}\bar{G}^{\left( 2\right)
c_{1}[i_{1}i_{2}}A^{c_{2}i_{3}}\cdots A^{c_{n-2}i_{n-1}]}\right) ,
\label{fpt29}
\end{eqnarray}
\begin{equation}
\left[ \tilde{G}_{a}^{(2)},\bar{G}_{ij}^{(2)b}\right] =g\left( \frac{%
\partial W_{ac}}{\partial \varphi _{b}}\bar{G}_{ij}^{(2)c}-\frac{\partial
^{2}W_{ac}}{\partial \varphi _{b}\partial \varphi _{d}}\bar{\gamma}%
_{d[i}^{(2)}A_{j]}^{c}\right) ,  \label{fpt30}
\end{equation}
\begin{equation}
\left[ \tilde{G}_{a}^{(2)},\bar{\gamma}_{b}^{(2)i}\right] =-g\frac{\partial
W_{ab}}{\partial \varphi _{c}}\bar{\gamma}_{c}^{(2)i}.  \label{fpt31}
\end{equation}
If we compare the expressions (\ref{fpt26})--(\ref{fpt31}) with the similar
results from \cite{noi}, we observe that here appear some supplementary
contributions, due to the presence of the terms from the deformed BRST
charge that break the PT invariance. Actually, from (\ref{fpt16'}) one can
read the entire tensor gauge structure of the first-class constraints by
analyzing the various polynomials in ghosts and antighosts. For instance,
the relations (\ref{fpt29})--(\ref{fpt31}) signify that the gauge algebra of
the first-class constraints is open (only closes on the first-class
constraint surface) and, meanwhile, offer us the concrete form of the
first-order structure functions. Still, higher-order structure functions
appear by taking repeatedly the Dirac brackets among more than two deformed
first-class constraint functions and can be read from the corresponding
polynomials of higher antighost number in (\ref{fpt16'}). Apart from
exhibiting an intricate gauge algebra, the deformed first-class constraints
remain reducible of order $\left( D-2\right) $, like those corresponding to
the free model. The structure of the reducibility functions and relations is
completely revealed by some of the terms of antighost number greater or
equal to one from (\ref{fpt16'}). These pieces are not modified by the
presence of the terms that break the PT invariance, being the same like in
\cite{noi}. However, for the sake of completeness, they are discussed in the
Appendix \ref{red}.

Let us analyze now the deformed BRST-invariant Hamiltonian (\ref{h6}). Its
component of antighost number zero,
\begin{equation}
\hat{H}_{0}=\int d^{n-1}x\left( -H_{i}^{a}\bar{\gamma}_{a}^{(2)i}+\frac{1}{2}%
B_{a}^{ij}\bar{G}_{ij}^{(2)a}+A_{0}^{a}\tilde{G}_{a}^{(2)}\right) ,
\label{fpt32}
\end{equation}
is nothing but the first-class Hamiltonian of the coupled model. From the
terms of antighost number zero we determine the deformed Dirac brackets
between the new first-class constraints and the first-class Hamiltonian of
the interacting theory under the form
\begin{equation}
\left[ \hat{H}_{0},G_{a}^{(1)}\right] =\tilde{G}_{a}^{(2)},  \label{fpt33}
\end{equation}
\begin{eqnarray}
\left[ \hat{H}_{0},\tilde{G}_{a}^{(2)}\right] &=&g\frac{\partial W_{ab}}{%
\partial \varphi _{c}}\left( A_{0}^{b}\tilde{G}_{c}^{(2)}-H_{i}^{b}\bar{%
\gamma}_{c}^{(2)i}-\frac{1}{2}\bar{G}_{ij}^{(2)b}B_{c}^{ij}\right)  \nonumber
\\
&&+g\frac{\partial ^{2}W_{ac}}{\partial \varphi _{b}\partial \varphi _{d}}%
\left( \frac{1}{2}B_{b}^{ij}\bar{\gamma}%
_{d[i}^{(2)}A_{j]}^{c}-B_{d0i}A_{0}^{c}\bar{\gamma}_{b}^{(2)i}\right)
\nonumber \\
&&-g\epsilon _{0i_{1}\ldots i_{n-1}}\left( N_{ab_{1}\ldots b_{n-1}}A^{b_{1}0}%
\bar{G}^{\left( 2\right) b_{2}[i_{1}i_{2}}A^{b_{3}i_{3}}\cdots
A^{b_{n-1}i_{n-1}]}\right.  \nonumber \\
&&\left. -\frac{\partial N_{ab_{1}\ldots b_{n-1}}}{\partial \varphi _{c}}%
A^{b_{1}0}\bar{\gamma}_{c}^{\left( 2 \right)
[i_{1}}A^{b_{2}i_{2}}\cdots A^{b_{n-1}i_{n-1}]}\right) ,
\label{fpt34}
\end{eqnarray}
\begin{equation}
\left[ \hat{H}_{0},G_{ij}^{(1)a}\right] =\bar{G}_{ij}^{(2)a},  \label{fpt35}
\end{equation}
\begin{equation}
\left[ H_{T},\gamma _{a}^{(1)i}\right] =\bar{\gamma}_{a}^{(2)i},
\label{fpt43}
\end{equation}
\begin{equation}
\left[ \hat{H}_{0},\bar{G}_{ij}^{(2)a}\right] =g\left( \frac{\partial W_{bc}%
}{\partial \varphi _{a}}A_{0}^{b}\bar{G}_{ij}^{(2)c}-\frac{\partial
^{2}W_{cd}}{\partial \varphi _{a}\partial \varphi _{b}}A_{0}^{c}\bar{\gamma}%
_{b[i}^{(2)}A_{j]}^{d}\right) ,  \label{fpt36}
\end{equation}
\begin{equation}
\left[ \hat{H}_{0},\bar{\gamma}_{a}^{(2)i}\right] =g\frac{\partial W_{ab}}{%
\partial \varphi _{c}}A_{0}^{b}\bar{\gamma}_{c}^{(2)i}.  \label{fpt37}
\end{equation}
It is simple to see that the relations (\ref{fpt32})--(\ref{fpt37}) also
contain nontrivial contributions due to the terms from the deformed
BRST-invariant Hamiltonian that break the PT invariance. The other terms in (%
\ref{h6}) reveal a new set of hamiltonian structure functions,
that follow by taking repeatedly the Dirac brackets involving the
interacting first-class Hamiltonian and more than two deformed
first-class constraint functions. This sort of structure
functions, as well as the equations satisfied by them, can be
written down directly from (\ref{h6}) by identifying the suitable
polynomials in ghosts and antighosts.

If we pass to the lagrangian formulation of the interacting theory (via the
extended and total actions, together with the accompanying gauge
symmetries), then we obtain that the interacting model is described by the
lagrangian action
\begin{eqnarray}
S\left[ A_{\mu }^{a},H_{\mu }^{a},\varphi _{a},B_{a}^{\mu \nu }\right]
&=&\int d^{n}x\left( H_{\mu }^{a}D^{\mu }\varphi _{a}+\frac{1}{2}B_{a}^{\mu
\nu }\bar{F}_{\mu \nu }^{a}\right.  \nonumber \\
&&\left. -\frac{1}{n}g\epsilon _{\mu _{1}\ldots \mu _{n}}N_{a_{1}\ldots
a_{n}}A^{a_{1}\mu _{1}}\cdots A^{a_{n}\mu _{n}}\right) ,  \label{fpt38}
\end{eqnarray}
invariant under the gauge transformations
\begin{equation}
\delta _{\epsilon }A_{\mu }^{a}=\left( D_{\mu }\right) _{\;\;b}^{a}\epsilon
^{b},  \label{fpt39}
\end{equation}
\begin{equation}
\delta _{\epsilon }\varphi _{a}=-gW_{ab}\epsilon ^{b},  \label{fpt40}
\end{equation}
\begin{eqnarray}
\delta _{\epsilon }H_{\mu }^{a} &=&\left( D^{\nu }\right)
_{\;\;b}^{a}\epsilon _{\mu \nu }^{b}-g\frac{\partial W_{bc}}{\partial
\varphi _{a}}\epsilon ^{b}H_{\mu }^{c}  \nonumber \\
&&+g\frac{\partial ^{2}W_{cd}}{\partial \varphi _{a}\partial \varphi _{b}}%
\left( \frac{1}{2}A^{c\nu }A^{d\rho }\epsilon _{b\mu \nu \rho }+A^{d\nu
}\epsilon ^{c}B_{b\mu \nu }\right)  \nonumber \\
&&-g\epsilon _{\mu \mu _{1}\ldots \mu _{n-1}}\frac{\partial N_{bc_{1}\ldots
c_{n-1}}}{\partial \varphi _{a}}\epsilon ^{b}A^{c_{1}\mu _{1}}\cdots
A^{c_{n-1}\mu _{n-1}},  \label{fpt41}
\end{eqnarray}
\begin{eqnarray}
\delta _{\epsilon }B_{a}^{\mu \nu } &=&\left( D_{\rho }\right)
_{a}^{\;\;b}\epsilon _{b}^{\mu \nu \rho }+gW_{ab}\epsilon ^{b\mu \nu }-g%
\frac{\partial W_{ab}}{\partial \varphi _{c}}\epsilon ^{b}B_{c}^{\mu \nu }
\nonumber \\
&&+g\left( n-1\right) g^{\mu \mu _{1}}g^{\nu \mu _{2}}\epsilon _{\mu
_{1}\ldots \mu _{n}}N_{abc_{1}\ldots c_{n-2}}\epsilon ^{b}A^{c_{1}\mu
_{3}}\cdots A^{c_{n-2}\mu _{n}}.  \label{fpt42}
\end{eqnarray}
At this stage it is clear that the deformation of the lagrangian gauge
transformations is a consequence of the deformation of the first-class
constraints like in (\ref{fpt26})--(\ref{fpt28}). The above gauge
transformations are on-shell $\left( n-2\right) $-order reducible, i.e., the
reducibility relations only hold on the stationary surface of the field
equations for the coupled action (\ref{fpt38}), while the accompanying gauge
algebra is open (the commutators among the deformed gauge transformations
only close on-shell). It is interesting to notice that the lagrangian
formulation of the interacting BF theory contains contributions that break
the PT invariance.

\section{Conclusion}

In conclusion in this paper we have investigated the PT-symmetry
breaking, consistent hamiltonian interactions in all $n\geq 4$
spacetime dimensions that can be added to an abelian BF model
involving a set of scalar fields, two sorts of one-forms, and a
system of two-forms by means of the hamiltonian deformation
procedure based on local BRST cohomology. The results related to
the deformation of the BRST charge can be synthesized by the fact
that only the first-order deformation is nontrivial, while its
consistency relies on the existence of a Poisson two-tensor on a
target space parametrized by the scalar fields. The deformation of
the BRST-invariant Hamiltonian stops also at order one in the
coupling constant and proves to be exact with respect to the
deformed BRST charge, which confirms the topological character of
the resulting interacting model. Both deformed BRST ingredients
contain terms that are not PT invariant. The associated coupled
theory is an interacting, topological BF model exhibiting an open
gauge algebra (the Dirac brackets among the deformed first-class
constraint functions only close on the first-class surface) and
on-shell reducibility (the reducibility relations take place on
the first-class surface).

\section*{Acknowledgment}

The authors are partially supported by the European Commission FP6 program
MRTN-CT-2004-005104 and by the grant A304/2004 with the Romanian National
Council for Academic Scientific Research (C.N.C.S.I.S.) and the Romanian
Ministry of Education and Research (M.E.C.).

\appendix

\section{Deformed BRST charge with PT invariance\label{ptinv}}

In this Appendix we briefly recall the structure of the component from the
deformed BRST charge (\ref{xc1}) that preserves the PT invariance, namely, $%
\omega _{1}$. Various details on the expression of $\omega _{1}$ and on the
interpretation of its various terms can be found in \cite{noi}. As it has
been mentioned in Section \ref{defch}, $\omega _{1}$ is subject to the
equation
\begin{equation}
s\omega _{1}=\partial _{i}j^{i},  \label{om1}
\end{equation}
for some local $j^{i}$. Taking into account the general cohomological
properties of the free BF model described by (\ref{f1}) and discussed in the
subsection \ref{firordch}, we can state that the most general decomposition
of $\omega _{1}$ along the antighost number can be taken to stop in
antighost number $\left( n-1\right) $%
\begin{equation}
\omega _{1}=\sum_{k=1}^{n-1}\stackrel{\left( k\right) }{\omega }_{1}.
\label{qw1}
\end{equation}
Furthermore, the last component, $\stackrel{(n-1)}{\omega }_{1}$, belongs to
the space of pure ghost number equal to $n$ from the cohomology of the
exterior longitudinal derivative, $\gamma \stackrel{(n-1)}{\omega }_{1}=0$,
and thus it is of the form (\ref{c5}), with the corresponding `invariant
polynomial' from $H_{n-1}^{\mathrm{inv}}\left( \delta |\tilde{d}\right) $,
and hence of the type (\ref{c9}). Requiring in addition that $\stackrel{(n-1)%
}{\omega }_{1}$ preserves the PT invariance, it follows that we can write
\begin{equation}
\stackrel{(n-1)}{\omega }_{1}=-W_{ab}^{i_{1}\ldots i_{n-1}}\eta
^{a}C_{i_{1}\ldots i_{n-1}}^{b}-\frac{\left( -\right) ^{n}}{2}\left(
M_{ab}^{c}\right) ^{i_{1}\ldots i_{n-1}}\eta ^{a}\eta ^{b}\eta
_{ci_{1}\ldots i_{n-1}},  \label{pt14}
\end{equation}
where
\begin{eqnarray}
W_{ab}^{i_{1}\ldots i_{n-1}} &=&\frac{\partial W_{ab}}{\partial \varphi _{c}}%
P_{c}^{i_{1}\ldots i_{n-1}}+\sum\limits_{p=2}^{n-1}\sum\limits_{1\leq
j_{1}\leq j_{2}\leq \cdots \leq j_{p}<n-1}\frac{\partial ^{p}W_{ab}}{%
\partial \varphi _{a_{1}}\partial \varphi _{a_{2}}\cdots \partial \varphi
_{a_{p}}}\times  \nonumber \\
&&\times P_{a_{1}}^{[i_{1}\ldots i_{j_{1}}}P_{a_{2}}^{i_{j_{1}+1}\ldots
i_{j_{1}+j_{2}}}\cdots P_{a_{p-1}}^{i_{j_{1}+\cdots +.j_{p-2}+1}\ldots
i_{j_{1}+\cdots +.j_{p-1}}}P_{a_{p}}^{i_{j_{1}+\cdots +.j_{p-1}+1}\ldots
i_{n-1}]}  \label{pt15}
\end{eqnarray}
and
\begin{eqnarray}
\left( M_{ab}^{c}\right) ^{i_{1}\ldots i_{n-1}} &=&\frac{\partial M_{ab}^{c}%
}{\partial \varphi _{d}}P_{d}^{i_{1}\ldots
i_{n-1}}+\sum\limits_{p=2}^{n-1}\sum\limits_{1\leq j_{1}\leq j_{2}\leq
\cdots \leq j_{p}<n-1}\frac{\partial ^{p}M_{ab}^{c}}{\partial \varphi
_{a_{1}}\partial \varphi _{a_{2}}\cdots \partial \varphi _{a_{p}}}\times
\nonumber \\
&&\times P_{a_{1}}^{[i_{1}\ldots i_{j_{1}}}P_{a_{2}}^{i_{j_{1}+1}\ldots
i_{j_{1}+j_{2}}}\cdots P_{a_{p-1}}^{i_{j_{1}+\cdots +j_{p-2}+1}\ldots
i_{j_{1}+\cdots +j_{p-1}}}P_{a_{p}}^{i_{j_{1}+\cdots +j_{p-1}+1}\ldots
i_{n-1}]},  \label{pt16}
\end{eqnarray}
with $W_{ab}$ and $M_{ab}^{c}=-M_{ab}^{c}$ depending only on the
undifferentiated scalar fields. The term of antighost number $\left(
n-2\right) $ from $\omega _{1}$ is subject to the equation $\delta \stackrel{%
(n-1)}{\omega }_{1}+\gamma \stackrel{(n-2)}{\omega }_{1}=\partial _{i}%
\stackrel{(n-2)}{m}^{i}$ and is consequently given by
\begin{eqnarray}
\stackrel{(n-2)}{\omega }_{1} &=&-W_{ab}^{i_{1}\ldots i_{n-2}}\eta
^{a}C_{i_{1}\ldots i_{n-2}}^{b}+\frac{\left( -\right) ^{n}}{2}\left(
M_{ab}^{c}\right) ^{i_{1}\ldots i_{n-2}}\eta ^{a}\eta ^{b}\eta
_{ci_{1}\ldots i_{n-2}}  \nonumber \\
&&-C_{n-1}^{1}W_{ab}^{i_{1}\ldots i_{n-2}}A^{ai_{n-1}}C_{i_{1}\ldots
i_{n-1}}^{b}  \nonumber \\
&&-\sum\limits_{k=3}^{n}\left( -\right) ^{k}C_{n-1}^{k-1}W_{ab}^{i_{1}\ldots
i_{n-k}}\mathcal{P}^{ai_{n-k+1}\ldots i_{n-1}}C_{i_{1}\ldots i_{n-1}}^{b}
\nonumber \\
&&-\left( -\right) ^{n}C_{n-1}^{1}\left( M_{ab}^{c}\right) ^{i_{1}\ldots
i_{n-2}}A^{ai_{n-1}}\eta ^{b}\eta _{ci_{1}\ldots i_{n-1}}  \nonumber \\
&&-\left( -\right) ^{n}\sum\limits_{k=3}^{n}\left( -\right)
^{k}C_{n-1}^{k-1}\left( M_{ab}^{c}\right) ^{i_{1}\ldots i_{n-k}}\mathcal{P}%
^{ai_{n-k+1}\ldots i_{n-1}}\eta ^{b}\eta _{ci_{1}\ldots i_{n-1}},
\label{pt20}
\end{eqnarray}
where $C_{n}^{k}$ denotes the number of combinations of $k$ objects drawn
from $n$, while the elements $W_{ab}^{i_{1}\ldots i_{n-2}}$ and $\left(
M_{ab}^{c}\right) ^{i_{1}\ldots i_{n-2}}$ result from (\ref{pt15})--(\ref
{pt16}) where we make the replacement $n\rightarrow n-1$. The component of
antighost number $\left( n-3\right) $ is solution to the equation $\delta
\stackrel{(n-2)}{\omega }_{1}+\gamma \stackrel{(n-3)}{\omega }_{1}=\partial
_{i}\stackrel{(n-3)}{m}^{i}$ and reads as
\begin{eqnarray}
\stackrel{(n-3)}{\omega }_{1} &=&-W_{ab}^{i_{1}i_{2}\ldots i_{n-3}}\eta
^{a}C_{i_{1}i_{2}\ldots i_{n-3}}^{b}  \nonumber \\
&&-\frac{\left( -\right) ^{n}}{2}\left( M_{ab}^{c}\right) ^{i_{1}i_{2}\ldots
i_{n-3}}\eta ^{a}\eta ^{b}\eta _{ci_{1}i_{2}\ldots i_{n-3}}  \nonumber \\
&&-C_{n-2}^{1}W_{ab}^{i_{1}i_{2}\ldots
i_{n-3}}A^{ai_{n-2}}C_{i_{1}i_{2}\ldots i_{n-2}}^{b}  \nonumber \\
&&+\sum\limits_{k=4}^{n}\left( -\right)
^{k}C_{n-2}^{k-2}W_{ab}^{i_{1}i_{2}\ldots i_{n-k}}\mathcal{P}%
^{ai_{n-k+1}\ldots i_{n-2}}C_{i_{1}i_{2}\ldots i_{n-2}}^{b}  \nonumber \\
&&+\left( -\right) ^{n}C_{n-2}^{1}\left( M_{ab}^{c}\right)
^{i_{1}i_{2}\ldots i_{n-3}}A^{ai_{n-2}}\eta ^{b}\eta _{ci_{1}i_{2}\ldots
i_{n-2}}  \nonumber \\
&&-\left( -\right) ^{n}\sum\limits_{k=4}^{n}\left( -\right)
^{k}C_{n-2}^{k-2}\left( M_{ab}^{c}\right) ^{i_{1}i_{2}\ldots i_{n-k}}%
\mathcal{P}^{ai_{n-k+1}\ldots i_{n-2}}\eta ^{b}\eta _{ci_{1}i_{2}\ldots
i_{n-2}}  \nonumber \\
&&-\left( -\right) ^{n}\sum\limits_{p=2}^{\left[ \frac{n-2}{2}\right]
}\sum\limits_{q=p+1}^{n-p-1}\left( -\right)
^{q}C_{n-1}^{p}C_{n-p-1}^{q}\left( M_{ab}^{c}\right) ^{i_{1}\ldots
i_{n-p-q-1}}\times  \nonumber \\
&&\times \mathcal{P}^{aj_{1}\ldots j_{q}}\mathcal{P}^{bl_{1}\ldots
l_{p}}\eta _{ci_{1}\ldots i_{n-p-q-1}j_{1}\ldots j_{q}l_{1}\ldots l_{p}}
\nonumber \\
&&-\frac{\left( -\right) ^{n}}{2}\sum\limits_{k=2}^{\left[ \frac{n-1}{2}%
\right] }\left( -\right) ^{k}C_{n-1}^{k}C_{n-k-1}^{k}\left(
M_{ab}^{c}\right) ^{i_{1}\ldots i_{n-2k-1}}\times  \nonumber \\
&&\times \mathcal{P}^{aj_{1}\ldots j_{k}}\mathcal{P}^{bl_{1}\ldots
l_{k}}\eta _{ci_{1}\ldots i_{n-2k-1}j_{1}\ldots j_{k}l_{1}\ldots l_{k}}
\nonumber \\
&&+\left( -\right) ^{n}C_{n-1}^{2}\left( M_{ab}^{c}\right) ^{i_{1}\ldots
i_{n-3}}A^{ai_{n-2}}A^{bi_{n-1}}\eta _{ci_{1}i_{2}\ldots i_{n-1}},
\label{pt21}
\end{eqnarray}
where $W_{ab}^{i_{1}i_{2}\ldots i_{n-3}}$ and $\left( M_{ab}^{c}\right)
^{i_{1}i_{2}\ldots i_{n-3}}$ are obtained from (\ref{pt15})--(\ref{pt16})
via the shift $n\rightarrow n-2$. Along the same line, we deduce that the
pieces of antighost number $\left( n-m\right) $, for $m=\overline{4,n-2}$,
are given by
\begin{eqnarray}
\stackrel{(n-m)}{\omega }_{1} &=&-W_{ab}^{i_{1}i_{2}\ldots i_{n-m}}\eta
^{a}C_{i_{1}i_{2}\ldots i_{n-m}}^{b}  \nonumber \\
&&-C_{n-m+1}^{1}W_{ab}^{i_{1}i_{2}\ldots
i_{n-m}}A^{ai_{n-m+1}}C_{i_{1}i_{2}\ldots i_{n-m+1}}^{b}  \nonumber \\
&&-\sum\limits_{k=m+1}^{n}\left( -\right)
^{k+m}C_{n-m+1}^{k-m+1}W_{ab}^{i_{1}i_{2}\ldots i_{n-k}}\mathcal{P}%
^{ai_{n-k+1}\ldots i_{n-m+1}}C_{i_{1}i_{2}\ldots i_{n-m+1}}^{b}  \nonumber \\
&&+\frac{\left( -\right) ^{m+n}}{2}\left( M_{ab}^{c}\right) ^{i_{1}\ldots
i_{n-m}}\eta ^{a}\eta ^{b}\eta _{ci_{1}i_{2}\ldots i_{n-m}}  \nonumber \\
&&-\left( -\right) ^{m+n}C_{n-m+1}^{1}\left( M_{ab}^{c}\right) ^{i_{1}\ldots
i_{n-m}}A^{ai_{n-m+1}}\eta ^{b}\eta _{ci_{1}i_{2}\ldots i_{n-m+1}}  \nonumber
\\
&&-\sum\limits_{k=m+1}^{n}\left( -\right) ^{k+n}C_{n-m+1}^{k-m+1}\left(
M_{ab}^{c}\right) ^{i_{1}\ldots i_{n-k}}\mathcal{P}^{ai_{n-k+1}\ldots
i_{n-m+1}}\eta ^{b}\eta _{ci_{1}i_{2}\ldots i_{n-m+1}}  \nonumber \\
&&-\left( -\right) ^{m+n}C_{n-m+2}^{2}\left( M_{ab}^{c}\right) ^{i_{1}\ldots
i_{n-m}}A^{ai_{n-m+1}}A^{bi_{n-m+2}}\eta _{ci_{1}i_{2}\ldots i_{n-m+2}}
\nonumber \\
&&+\frac{1}{2}\sum\limits_{k=2}^{\left[ \frac{n-m+2}{2}\right] }\left(
-\right) ^{k+n+m}C_{n-m+2}^{k}C_{n-m-k+2}^{k}\left( M_{ab}^{c}\right)
^{i_{1}\ldots i_{n-2k-m+2}}\times  \nonumber \\
&&\times \mathcal{P}^{aj_{1}\ldots j_{k}}\mathcal{P}^{bl_{1}\ldots
l_{k}}\eta _{ci_{1}\ldots i_{n-2k-m+2}j_{1}\ldots j_{k}l_{1}\ldots l_{k}}
\nonumber \\
&&+\sum\limits_{p=2}^{\left[ \frac{n-m+1}{2}\right]
}\sum\limits_{q=p+1}^{n-m-p+2}\left( -\right)
^{q+n+m}C_{n-m+2}^{p}C_{n-m-p+2}^{q}\left( M_{ab}^{c}\right)
^{i_{1}\ldots
i_{n-m-p-q+2}}\times  \nonumber \\
&&\times \mathcal{P}^{aj_{1}\ldots j_{q}}\mathcal{P}^{bl_{1}\ldots
l_{p}}\eta _{ci_{1}\ldots i_{n-m-p-q+2}j_{1}\ldots j_{q}l_{1}\ldots l_{p}}
\nonumber \\
&&-\sum\limits_{k=m+1}^{n}\left( -\right)
^{k+n}C_{n-m+1}^{k-m+1}C_{n-m+2}^{1}\left( M_{ab}^{c}\right) ^{i_{1}\ldots
i_{n-k}}\times  \nonumber \\
&&\times \mathcal{P}^{ai_{n-k+1}\ldots i_{n-m+1}}A^{bi_{n-m+2}}\eta
_{ci_{1}i_{2}\ldots i_{n-m+2}},  \label{pt22}
\end{eqnarray}
where the elements $W_{ab}^{i_{1}i_{2}\ldots i_{n-m}}$ and $\left(
M_{ab}^{c}\right) ^{i_{1}\ldots i_{n-m}}$ are deduced from (\ref{pt15})--(%
\ref{pt16}) where we set $n\rightarrow n-m+1$. Finally, the components of
antighost number one and zero from the first-order deformation $\omega _{1}$
respectively take the form
\begin{eqnarray}
\stackrel{(1)}{\omega }_{1} &=&-\frac{\partial W_{ab}}{\partial \varphi _{c}}%
P_{c}^{i}\left( \eta ^{a}C_{i}^{b}+2A^{aj}C_{ij}^{b}\right) +W_{ab}\mathcal{P%
}^{aij}C_{ij}^{b}  \nonumber \\
&&+\frac{\partial M_{ab}^{c}}{\partial \varphi _{d}}P_{di}\left( \frac{1}{2}%
\eta ^{a}\eta ^{b}B_{c}^{0i}+2A_{j}^{a}\eta ^{b}\eta
_{c}^{ij}+3A_{j}^{a}A_{k}^{b}\eta _{c}^{ijk}\right)  \nonumber \\
&&-M_{ab}^{c}\left( \mathcal{P}_{ij}^{a}\eta ^{b}\eta _{c}^{ij}+\mathcal{P}%
_{[ij}^{a}A_{k]}^{b}\eta _{c}^{ijk}+\frac{1}{2}\eta ^{a}\eta ^{b}\mathcal{P}%
_{c}\right) ,  \label{pt23}
\end{eqnarray}
\begin{equation}
\stackrel{(0)}{\omega }_{1}=W_{ab}\left( \eta
^{a}H_{0}^{b}-A^{ai}C_{i}^{b}\right) +M_{ab}^{c}\left( A_{i}^{a}\eta
^{b}B_{c}^{0i}-A_{i}^{a}A_{j}^{b}\eta _{c}^{ij}\right) .  \label{pt24}
\end{equation}
In conclusion, the first-order deformation of the BRST charge for
the model under study that preserves the PT invariance is the sum
among the components (\ref{pt14}) and (\ref {pt20})--(\ref{pt24}).

\section{Reducibility of the deformed interacting model\label{red}}

Here, we investigate the reducibility of the first-class constraints
corresponding to the interacting model discussed in Section \ref{int}. In
view of this, we analyze some of the terms contained in the deformed BRST
charge given in (\ref{fpt16'}), with the corresponding components listed in (%
\ref{fpt2}), (\ref{fpt4})--(\ref{fpt6}), (\ref{fpt7})--(\ref{fpt8}) and
respectively in (\ref{pt14}), (\ref{pt20})--(\ref{pt24}). From the elements
in (\ref{fpt16'}) with antighost number one that are linear in the ghosts
one reads the first-order reducibility relations
\begin{eqnarray}
\left( \bar{Z}_{i_{1}i_{2}i_{3}}^{a}\right) _{b}^{ij}\bar{G}%
_{ij}^{(2)b}+\left( \bar{Z}_{i_{1}i_{2}i_{3}}^{a}\right) _{i}^{b}\bar{\gamma}%
_{b}^{(2)i} &=&0,  \label{pt53} \\
\left( \bar{Z}_{a}^{i_{1}i_{2}}\right) _{b}^{ij}\bar{G}_{ij}^{(2)b}+\left(
\bar{Z}_{a}^{i_{1}i_{2}}\right) _{i}^{b}\bar{\gamma}_{b}^{(2)i} &=&0,
\label{pt54}
\end{eqnarray}
and also the associated first-order reducibility functions
\begin{eqnarray}
\left( \bar{Z}_{i_{1}i_{2}i_{3}}^{a}\right) _{b}^{ij} &=&\frac{1}{2}\left(
D_{[i_{1}}\right) _{\;\;b}^{a}\delta _{i_{2}}^{i}\delta _{i_{3}]}^{j},
\label{pt55} \\
\left( \bar{Z}_{i_{1}i_{2}i_{3}}^{a}\right) _{i}^{b} &=&g\frac{\partial
^{2}W_{cd}}{\partial \varphi _{a}\partial \varphi _{b}}%
g_{i[i_{1}}A_{i_{2}}^{c}A_{i_{3}]}^{d},  \label{pt56} \\
\left( \bar{Z}_{a}^{i_{1}i_{2}}\right) _{b}^{ij} &=&-\frac{1}{2}%
gW_{ab}\left( g^{i_{1}i}g^{i_{2}j}-g^{i_{1}j}g^{i_{2}i}\right) ,
\label{pt57} \\
\left( \bar{Z}_{a}^{i_{1}i_{2}}\right) _{i}^{b} &=&-\left( D^{[i_{1}}\right)
_{a}^{\;\;b}\delta _{i}^{i_{2}]},  \label{pt58}
\end{eqnarray}
where
\begin{equation}
\left( D_{i}\right) _{\;\;b}^{a}=\delta _{b}^{a}\partial _{i}-g\frac{%
\partial W_{bc}}{\partial \varphi _{a}}A_{i}^{c}.  \label{pt59}
\end{equation}
The part from (\ref{fpt16'}) that is linear in the ghosts with the pure
ghost number equal to $k+1\geq 3$ contains polynomials of antighost number $%
k\geq 2$, which are at least quadratic in the antighosts, so the
reducibility relations of order $k\geq 2$ only close on the
first-class
constraint surface (on-shell). For instance, in pure ghost number three ($%
k+1=3$) the second-order reducibility relations take place on-shell
\begin{eqnarray}
&&\left( \bar{Z}_{i_{1}i_{2}i_{3}i_{4}}^{a}\right)
_{b}^{j_{1}j_{2}j_{3}}\left( \bar{Z}_{j_{1}j_{2}j_{3}}^{b}\right)
_{c}^{ij}f_{ij}^{c}+\left(
\bar{Z}_{i_{1}i_{2}i_{3}i_{4}}^{a}\right) _{j_{1}j_{2}}^{b}\left(
\bar{Z}_{b}^{j_{1}j_{2}}\right) _{c}^{ij}f_{ij}^{c}
\nonumber \\
&&=-g\left( \frac{\partial W_{bc}}{\partial \varphi _{a}}\bar{G}%
_{[i_{1}i_{2}}^{(2)b}f_{i_{3}i_{4}]}^{c}-\frac{\partial ^{2}W_{cd}}{\partial
\varphi _{a}\partial \varphi _{b}}\bar{\gamma}%
_{b[i_{1}}^{(2)}A_{i_{2}}^{c}f_{i_{3}i_{4}]}^{d}\right) ,  \label{pt60}
\end{eqnarray}
\begin{eqnarray}
&&\left( \bar{Z}_{a}^{i_{1}i_{2}i_{3}}\right) _{j_{1}j_{2}}^{b}\left( \bar{Z}%
_{b}^{j_{1}j_{2}}\right) _{i}^{c}f_{c}^{i}+\left( \bar{Z}%
_{a}^{i_{1}i_{2}i_{3}}\right) _{b}^{j_{1}j_{2}j_{3}}\left( \bar{Z}%
_{j_{1}j_{2}j_{3}}^{b}\right) _{i}^{c}f_{c}^{i} \nonumber \\
&&=g\left( \frac{\partial W_{ab}}{\partial \varphi _{c}}\bar{G}%
^{(2)b[i_{1}i_{2}}f_{c}^{i_{3}]}-\frac{\partial ^{2}W_{ab}}{\partial \varphi
_{c}\partial \varphi _{d}}\bar{\gamma}%
_{c}^{(2)[i_{1}}A^{bi_{2}}f_{d}^{i_{3}]}\right) ,  \label{pt61}
\end{eqnarray}
where $f_{c}^{i}$ and $f_{ij}^{c}$ are some arbitrary, smooth functions (the
latter ones are antisymmetric in their spatial indices). The reducibility
functions involved in (\ref{pt60}) and (\ref{pt61}) take the form
\begin{eqnarray}
\left( \bar{Z}_{i_{1}i_{2}i_{3}i_{4}}^{a}\right) _{b}^{j_{1}j_{2}j_{3}} &=&-%
\frac{1}{3!}\left( D_{[i_{1}}\right) _{\;\;b}^{a}\delta
_{i_{2}}^{j_{1}}\delta _{i_{3}}^{j_{2}}\delta _{i_{4}]}^{j_{3}},
\label{pt62} \\
\left( \bar{Z}_{i_{1}i_{2}i_{3}i_{4}}^{a}\right) _{j_{1}j_{2}}^{b} &=&-\frac{%
g}{2}g_{j_{1}k_{1}}g_{j_{2}k_{2}}\frac{\partial ^{2}W_{cd}}{\partial \varphi
_{a}\partial \varphi _{b}}\delta _{[i_{1}}^{k_{1}}\delta
_{i_{2}}^{k_{2}}A_{i_{3}}^{c}A_{i_{4}]}^{d},  \label{pt63} \\
\left( \bar{Z}_{a}^{i_{1}i_{2}i_{3}}\right) _{j_{1}j_{2}}^{b} &=&\frac{1}{2}%
\left( D^{[i_{1}}\right) _{a}^{\;\;b}\delta _{j_{1}}^{i_{2}}\delta
_{j_{2}}^{i_{3}]},  \label{pt64} \\
\left( \bar{Z}_{a}^{i_{1}i_{2}i_{3}}\right) _{b}^{j_{1}j_{2}j_{3}} &=&\frac{g%
}{3!}W_{ab}\sum\limits_{\sigma \in S_{3}}\left( -\right) ^{\sigma
}g^{i_{1}j_{\sigma (1)}}g^{i_{2}j_{\sigma (2)}}g^{i_{3}j_{\sigma (3)}}.
\label{pt65}
\end{eqnarray}
In (\ref{pt65}) $S_{3}$ is the set of permutations of the elements $%
\{1,2,3\}$ and $\left( -\right) ^{\sigma }$ denotes the parity of
the permutation $\sigma $ from $S_{3}$.

Reprising a similar analysis with respect to the terms from (\ref{fpt16'})
linear in the ghosts with the pure ghost number equal to $\left( p+1\right) $%
, $p=\overline{3,n-3}$, we deduce some reducibility relations of order $p$
that also close on-shell, namely,
\begin{eqnarray}
\left( \bar{Z}_{i_{1}\ldots i_{p+2}}^{a}\right) _{b}^{j_{1}\ldots
j_{p+1}}\left( \bar{Z}_{j_{1}\ldots j_{p+1}}^{b}\right) _{c}^{k_{1}\ldots
k_{p}}+\left( \bar{Z}_{i_{1}\ldots i_{p+2}}^{a}\right) _{j_{1}\ldots
j_{p}}^{b}\left( \bar{Z}_{b}^{j_{1}\ldots j_{p}}\right) _{c}^{k_{1}\ldots
k_{p}} &\approx &0,  \label{pt66} \\
\left( \bar{Z}_{a}^{i_{1}\ldots i_{p+1}}\right) _{j_{1}\ldots
j_{p}}^{b}\left( \bar{Z}_{b}^{j_{1}\ldots j_{p}}\right) _{k_{1}\ldots
k_{p-1}}^{c}+\left( \bar{Z}_{a}^{i_{1}\ldots i_{p+1}}\right)
_{b}^{j_{1}\ldots j_{p+1}}\left( \bar{Z}_{j_{1}\ldots j_{p+1}}^{b}\right)
_{k_{1}\ldots k_{p-1}}^{c} &\approx &0,  \label{pt67}
\end{eqnarray}
with the reducibility functions
\begin{eqnarray}
\left( \bar{Z}_{i_{1}\ldots i_{p+2}}^{a}\right) _{b}^{j_{1}\ldots j_{p+1}}
&=&\frac{\left( -\right) ^{p+1}}{\left( p+1\right) !}\left(
D_{[i_{1}}\right) _{\;\;b}^{a}\delta _{i_{2}}^{j_{1}}\cdots \delta
_{i_{p+2}]}^{j_{p+1}},  \label{pt68} \\
\left( \bar{Z}_{i_{1}\ldots i_{p+2}}^{a}\right) _{j_{1}\ldots j_{p}}^{b} &=&-%
\frac{\left( -\right) ^{p}g}{p!}g_{j_{1}k_{1}}\cdots g_{j_{p}k_{p}}\frac{%
\partial ^{2}W_{cd}}{\partial \varphi _{a}\partial \varphi _{b}}\delta
_{[i_{1}}^{k_{1}}\cdots \delta
_{i_{p}}^{k_{p}}A_{i_{p+1}}^{c}A_{i_{p+2}]}^{d},  \label{pt69} \\
\left( \bar{Z}_{a}^{i_{1}\ldots i_{p+1}}\right) _{j_{1}\ldots j_{p}}^{b} &=&%
\frac{\left( -\right) ^{p}}{p!}\left( D^{[i_{1}}\right) _{a}^{\;\;b}\delta
_{j_{1}}^{i_{2}}\cdots \delta _{j_{p}}^{i_{p+1}]},  \label{pt70} \\
\left( \bar{Z}_{a}^{i_{1}\ldots i_{p+1}}\right) _{b}^{j_{1}\ldots j_{p+1}}
&=&\frac{\left( -\right) ^{p}g}{\left( p+1\right) !}W_{ab}\sum\limits_{%
\sigma \in S_{p+1}}\left( -\right) ^{\sigma }g^{i_{1}j_{\sigma
(1)}}g^{i_{2}j_{\sigma (2)}}\cdots g^{i_{p+1}j_{\sigma (p+1)}}.  \label{pt71}
\end{eqnarray}
In (\ref{pt71}) $S_{p+1}$ is the set of permutations of the elements $%
\{1,\ldots ,p+1\}$, while $\left( -\right) ^{\sigma }$ stands for the parity
of the permutation $\sigma $ belonging to $S_{p+1}$. The reducibility
relations of maximum order, $\left( n-2\right) $, follow from the elements
in (\ref{fpt16'}) that are linear in the ghosts with the pure ghost number
equal to $\left( n-1\right) $
\begin{eqnarray}
&&\left( \bar{Z}_{a}^{i_{1}\ldots i_{n-1}}\right) _{j_{1}\ldots
j_{n-2}}^{b}\left( \bar{Z}_{b}^{j_{1}\ldots j_{n-2}}\right) _{k_{1}\ldots
k_{n-3}}^{c}f_{c}^{k_{1}\ldots k_{n-3}}  \nonumber \\
&&+\left( \bar{Z}_{a}^{i_{1}\ldots i_{n-1}}\right) _{b}^{j_{1}\ldots
j_{n-1}}\left( \bar{Z}_{j_{1}\ldots j_{n-1}}^{b}\right) _{k_{1}\ldots
k_{n-3}}^{c}f_{c}^{k_{1}\ldots k_{n-3}}  \nonumber \\
&=&-g\left( \frac{\partial ^{2}W_{ab}}{\partial \varphi _{c}\partial \varphi
_{d}}\bar{\gamma}_{c}^{(2)[i_{1}}A^{bi_{2}}f_{d}^{i_{3}\ldots i_{n-1}]}-%
\frac{\partial W_{ab}}{\partial \varphi _{c}}\bar{G}%
^{(2)b[i_{1}i_{2}}f_{c}^{i_{3}\ldots i_{n-1}]}\right) ,  \label{pt72}
\end{eqnarray}
\begin{eqnarray}
&&\left( \bar{Z}_{a}^{i_{1}\ldots i_{n-1}}\right) _{j_{1}\ldots
j_{n-2}}^{b}\left( \bar{Z}_{b}^{j_{1}\ldots j_{n-2}}\right)
_{c}^{k_{1}\ldots k_{n-2}}f_{k_{1}\ldots k_{n-2}}^{c}  \nonumber \\
&&+\left( \bar{Z}_{a}^{i_{1}\ldots i_{n-1}}\right) _{b}^{j_{1}\ldots
j_{n-1}}\left( \bar{Z}_{j_{1}\ldots j_{n-1}}^{b}\right) _{c}^{k_{1}\ldots
k_{n-2}}f_{k_{1}\ldots k_{n-2}}^{c}  \nonumber \\
&=&-g\frac{\partial W_{ab}}{\partial \varphi _{c}}\bar{\gamma}%
_{c}^{(2)[i_{1}}f^{bi_{2}\ldots i_{n-1}]},  \label{pt73}
\end{eqnarray}
with $f_{c}^{k_{1}\ldots k_{n-3}}$ and $f_{k_{1}\ldots k_{n-2}}^{c}$ some
arbitrary, smooth functions, completely antisymmetric in their spatial
indices. The reducibility functions of order $\left( n-2\right) $ are
consequently given by
\begin{eqnarray}
\left( \bar{Z}_{a}^{i_{1}\ldots i_{n-1}}\right) _{j_{1}\ldots j_{n-2}}^{b}
&=&\frac{\left( -\right) ^{n}}{\left( n-2\right) !}\left( D^{[i_{1}}\right)
_{a}^{\;\;b}\delta _{j_{1}}^{i_{2}}\cdots \delta _{j_{n-2}}^{i_{n-1}]},
\label{pt74} \\
\left( \bar{Z}_{a}^{i_{1}\ldots i_{n-1}}\right) _{b}^{j_{1}\ldots j_{n-1}}
&=&\frac{\left( -\right) ^{n}g}{\left( n-1\right) !}W_{ab}\sum\limits_{%
\sigma \in S_{n-1}}\left( -\right) ^{\sigma }g^{i_{1}j_{\sigma
(1)}}g^{i_{2}j_{\sigma (2)}}\cdots g^{i_{n-1}j_{\sigma (n-1)}}.  \label{pt75}
\end{eqnarray}
The notations $S_{n-1}$ and $\left( -\right) ^{\sigma }$ have the same
meanings like before.

\end{document}